\documentclass[5p,times,numbers,sort&compress]{elsarticle}  
\usepackage{graphicx}
\usepackage{booktabs}
\usepackage{amsmath,amssymb}
\usepackage{newtxtext,newtxmath} 

\usepackage{pgfgantt}
\usepackage{hyperref}
\usepackage{mathrsfs}
\usepackage{subcaption}
\usepackage{textcomp}
\usepackage{xcolor}
\usepackage{enumitem}
\usepackage{multirow}

\newproof{pf}{Proof}

\usepackage{algorithm}        
\usepackage{algpseudocode}    
\usepackage[most]{tcolorbox}  
\usepackage{caption}          
\algrenewcommand\algorithmicrequire{\textbf{Inputs:}}
\algrenewcommand\algorithmicensure{\textbf{Output:}}
\usepackage{stfloats}

\journal{Applied Energy}   

\begin{document}

\begin{frontmatter}

\title{Input Convex Encoder-Only Transformer for Computationally Efficient Model Predictive Control in Building Demand Response}

\author[1]{Kaipeng Xu}
\author[1]{Zhuo Zhi}
\author[1]{Keyue Jiang}

\address[1]{Department of Electronic and Electrical Engineering, University College London, Torrington Place, London WC1E 7JE, United Kingdom}


\begin{abstract}
Learning-based Model Predictive Control (MPC) has emerged as a powerful strategy for building demand response (DR). However, its practical deployment is often hindered by the non-convex optimization  problems induced by standard neural network models. These problems lead to long solver times and a lack of global optimality guarantees, making long-horizon real-time control challenging. Because forecasting building thermal dynamics and energy consumption relies heavily on historical time-series data, Input Convex Neural Networks (ICNNs) tailored for temporal tasks have been developed to address this issue. Notable examples include Input Convex Recurrent Neural Networks (ICRNNs) and Input Convex Long Short-Term Memory (IC-LSTM). Although IC-LSTM improves the modelling capability of simpler ICRNNs, its recurrent architecture remains computationally expensive and, in the experiments reported here, exhibits numerical training instability at longer historical sequence lengths. To address these limitations, this paper introduces the Input Convex Encoder-Only Transformer (IC-EoT), which combines an encoder-only attention architecture with input convexity guarantees. IC-EoT was evaluated in a co-simulation framework using the Energym Python library and the EnergyPlus building simulator, and compared with IC-LSTM and standard non-convex models in residential and commercial testbeds. IC-EoT maintained predictive accuracy comparable to IC-LSTM, showed no non-finite loss events across 60 multi-seed training runs, and reduced mean per-epoch training time by 26.2\% and 29.8\% in the two testbeds. At the eight-hour prediction horizon, IC-EoT reduced mean MPC solution time by factors of 5.8 and 5.3 relative to IC-LSTM, while providing comparable closed-loop cost and comfort performance.
\end{abstract}


\end{frontmatter}

\section{Introduction}
Model Predictive Control (MPC) is a widely used strategy for complex control tasks such as building demand response (DR), owing to its ability to handle constraints and optimize over a future horizon \cite{drgovna2020all, afroz2018modeling, knudsen2016demand, tang2019model}. Learning-based MPC, which uses machine learning (ML) models to capture system dynamics from data, offers a flexible and accurate alternative to traditional modelling approaches \cite{gonzalez2017data,azuatalam2020reinforcement,zhang2016optimal}. However, integrating standard ML models such as Recurrent Neural Networks (RNNs) or Transformers into MPC frameworks introduces a significant challenge: the resulting optimization problem becomes highly \textbf{non-convex} \cite{sankaranarayanan2022cdinn, pare2024efficient}. Solving such problems is computationally intensive, offers no guarantee of finding a global optimum, and is often too slow for the real-time decision-making required in many applications \cite{xing2024bilevel}. This often forces a trade-off, where simpler, less accurate linear models are used to maintain tractability \cite{wang2019data}, or where physics-based, first-principles models—those derived from fundamental physical laws like thermodynamics—are employed, which, while interpretable, are time-consuming to develop and require extensive domain expertise \cite{moriyasu2021structured}.

To address this trade-off between modelling accuracy and optimization tractability, Input Convex Neural Networks (ICNNs) were developed to guarantee convexity with respect to network inputs \cite{amos2017input}. By constraining activation functions to be convex and non-decreasing while restricting certain weights to be non-negative, ICNNs create a model class that permits efficient global optimization. Because accurately modelling complex systems such as buildings benefits from historical information on thermal states, control actions, and energy consumption, temporal input convex architectures have been developed to represent sequential dynamics more directly. This requirement motivated the development of temporal architectures, specifically Input Convex Recurrent Neural Networks (ICRNNs) \cite{chen2018optimal}, which paved the way for numerous successful applications. In the domain of building control, this framework has proven highly effective for achieving significant energy savings and reliable comfort maintenance \cite{chen2018optimal, bunning2021input, pare2024efficient}, with further benefits demonstrated in broader energy systems like chiller-pump systems and power units \cite{xing2024bilevel, zhu2022nonlinear}. To represent more complex temporal dynamics, Wang et al. proposed the Input Convex Long Short-Term Memory (IC-LSTM)~\cite{wang2025real}. IC-LSTM improves the representational capability of earlier recurrent ICNNs while preserving input convexity. However, its hidden and cell states must be propagated sequentially across the historical input window, and its recurrent update contains repeated multiplicative gate and state operations. These characteristics motivate the investigation of a non-recurrent input convex architecture and a systematic comparison of its computational cost and numerical training behaviour with those of the recurrent baseline.

To address this gap, this paper proposes the \textbf{Input Convex Encoder-Only Transformer (IC-EoT)}, an attention-based temporal ICNN that replaces recurrent state propagation with an encoder-only architecture. The model introduces a convex additive attention mechanism and convexity-preserving encoder components, while retaining the input convex structure required for tractable MPC. The study investigates whether this non-recurrent design can reduce computational cost and improve numerical training reliability relative to IC-LSTM without substantially degrading predictive or closed-loop control performance.

To validate the practical efficacy of the proposed IC-EoT, this study evaluates its performance in a challenging real-time control problem: building DR. Buildings represent one of the largest global energy consumers, with Heating, Ventilation, and Air Conditioning (HVAC) systems being a primary driver of this demand \cite{hu2017survey, center2021annual}. Their inherent thermal storage capacity, however, presents a significant opportunity for DR programs, allowing the building's thermal mass to act as a virtual battery \cite{jin2017user, hussain2015review}. By strategically pre-heating or pre-cooling during off-peak hours, energy consumption can be shifted away from high-price periods to reduce costs and support grid stability \cite{ma2012demand,vedullapalli2019combined}. Building DR therefore provides a relevant constrained-control application for evaluating the predictive, computational, and closed-loop properties of the proposed model.

The main contributions of this work are:
\begin{itemize}
    \item The development of IC-EoT, a non-recurrent temporal ICNN based on a convex additive attention mechanism and an encoder-only architecture.

    \item A formal analysis establishing the input convexity of the proposed architecture and the convexity of its recursively formulated soft-constrained MPC problem under the stated assumptions.

    \item An empirical comparison with IC-LSTM, the conventional Encoder-Only Transformer (EoT), and Long Short-Term Memory (LSTM) across residential and commercial building testbeds, covering predictive accuracy, numerical training stability, MPC solution time, solver termination behaviour, and closed-loop cost and comfort performance.
\end{itemize}

The remainder of this paper is organized as follows. Section \ref{sec:related_work} reviews the Transformer-based sequence modelling and the development of ICNNs, and identifies the architectural gap addressed in this work. Section \ref{sec:iceot} presents the IC-EoT architecture and establishes its input convex properties. Section \ref{sec:case_study} formulates the associated MPC problem, introduces the residential and commercial building testbeds, and evaluates predictive performance, numerical training stability, solver time, termination behaviour, and closed-loop control performance. Finally, Section \ref{sec:conclusion_future_work} summarises the key findings and outlines directions for future work.

\section{Related Work}
\label{sec:related_work}

The development of the proposed IC-EoT is situated at the intersection of two influential domains in ML: the attention-based \textbf{Transformer architecture} and the optimization-focused paradigm of \textbf{ICNNs}. This section first discusses the Transformer, focusing on its self-attention mechanism and parallel processing capabilities, which form the structural basis of our model. It then traces the progression of ICNNs, from their foundational feed-forward design to more complex recurrent structures, strictly detailing the formulations and convexity conditions established in the literature. Finally, this section culminates in identifying a critical research gap: the absence of an input convex model that leverages the parallel processing capabilities of the Transformer, a gap this work aims to fill.

\subsection{The Transformer and Self-Attention}
While the full design of the Transformer is detailed in the original work by Vaswani et al. \cite{vaswani2017attention}, this section focuses on its core features relevant to our study: the self-attention mechanism and parallel processing. The Transformer marked a significant departure from the traditional recurrent models used for sequence processing. Its core innovation lies in avoiding recurrence and convolutions, relying instead on a \textbf{self-attention mechanism} to model global dependencies between inputs and outputs. This design choice fundamentally allows for significantly more parallelization during training, drastically reducing the time required to train on large datasets~\cite{vaswani2017attention}. The effectiveness of this approach is further enhanced by \textbf{Multi-Head Attention}, which permits the model to jointly extract information from different representation subspaces. As the self-attention mechanism is permutation-invariant, the model injects \textbf{Positional Encodings} to make use of the sequence order~\cite{vaswani2017attention}. This combination of parallel processing and powerful dependency modelling has established the Transformer as the foundation for state-of-the-art performance in a vast array of tasks.

Since its introduction, researchers have actively modified the original self-attention mechanism to improve computational efficiency and adapt to specific tasks. Notable modifications include linear attention \cite{katharopoulos2020transformers}, additive attention \cite{wu2021fastformer, shaker2023swiftformer}, dynamic convolutions \cite{wu2019pay}, and auto-correlation mechanisms \cite{wu2021autoformer}. Beyond these specific algorithmic improvements, recent studies offer a more fundamental perspective on the Transformer's success. Specifically, the MetaFormer framework proposed by Yu et al. \cite{yu2022metaformer} argues that the general architecture itself is the primary contributor to the model's strong performance. The authors empirically demonstrated that replacing the complex self-attention module with a simple spatial pooling layer still yields highly competitive results. This critical insight directly inspires this research, providing the empirical confidence needed to fundamentally redesign the attention module. By replacing the standard non-convex scaled dot product attention with a guaranteed convex operator, this work leverages the structural advantages of the general Transformer backbone while strictly achieving the required input convexity.

\subsection{Progression of Input Convex Neural Networks}
While powerful, standard neural architectures like the Transformer produce highly non-convex input-output mappings, making their integration into optimization problems computationally challenging. To address this, the field of ICNNs emerged to create models that are convex by construction, thereby ensuring that downstream optimization problems are tractable and can be solved efficiently to a global optimum.

\subsubsection{Input Convex Feed-Forward Networks}
The foundational concept of ICNNs was introduced by Amos et al. \cite{amos2017input} with a feed-forward architecture. An L-layer Input Convex Feed-Forward Network (ICFNN) is defined by the recurrence:
\begin{equation}
    z_{l+1} = g_l(W_{l}^{(z)}z_l + W_{l}^{(x)}x + b_l), \quad l=0,1,...,L-1, \label{eq:icfnn}
\end{equation}
with $z_0 = 0$ and $W_0^{(z)} = 0$. Here, $x$ is the network input, $z_l$ is the hidden activation of the $l$-th layer, and $b_l$ represents the bias vector. The matrices $W_{l}^{(z)}$ are the weights connecting the hidden layers, while $W_{l}^{(x)}$ are "passthrough" weights connecting the original input directly to each hidden layer.

As established by Amos et al. \cite{amos2017input}, the output function is convex with respect to the input $x$ for single-step predictions if the hidden weights $W_{l}^{(z)}$ are \textbf{non-negative}, and the activation functions $g_l$ are \textbf{convex and non-decreasing}. The "passthrough" weights $W_{l}^{(x)}$ remain unconstrained, providing the network with essential representational flexibility \cite{amos2017input}.

However, when applying ICNNs to dynamical system control, Chen et al. \cite{chen2018optimal} identified a critical limitation. In MPC, multi-step predictions require composing the network with itself (i.e., the output state of one step becomes the input state for the next). To ensure the entire multi-step rollout maintains its convexity, they introduced a stricter variant. For multi-step continuous prediction, \textbf{all weights} mapping from previous states and inputs (i.e., both $W_{l}^{(z)}$ and $W_{l}^{(x)}$) must be \textbf{non-negative} \cite{chen2018optimal}.

Building directly upon this requirement, Bünning et al. \cite{bunning2021input} explicitly adapted the ICFNN architecture for building MPC. In their autoregressive formulation, the network input $x$ is partitioned into state variables and control actions. To guarantee a convex input-output relationship over an extended prediction horizon, they introduced targeted structural constraints to the original ICFNN: all weights corresponding to the \textbf{state variables} within the passthrough matrices $W_{l}^{(x)}$ must be strictly \textbf{non-negative} \cite{bunning2021input}. This adaptation ensures end-to-end multi-step convexity, making feed-forward ICNNs viable for real-time building control, albeit at the cost of further restricting the model's expressive capability.

\subsubsection{Input Convex Recurrent Neural Networks}
\label{Related_work: icrnn}
To directly model temporal dynamical systems, Chen et al. \cite{chen2018optimal} further extended the input convex property to the recurrent domain. The proposed ICRNN is defined by the state and output equations:
\begin{align}
    h_t &= g_1( U \hat{u}_t + W h_{t-1} + D_2 \hat{u}_{t-1} ), \label{eq:icrnn_h} \\
    y_t &= g_2( V h_t + D_1 h_{t-1} + D_3 \hat{u}_t ), \label{eq:icrnn_y}
\end{align}
where $u_t$ is the original input at time step $t$, $h_t$ is the hidden state, and $y_t$ is the final output. The matrices $U, W, V$ and $D_1, D_2, D_3$ are weight matrices that map the inputs and past states to the current state and output.

A crucial design element introduced here is the use of an \textbf{expanded input}, $\hat{u}_t = [u_t^\top, -u_t^\top]^\top$. Because the weight matrices mapping to the state and output must be non-negative to maintain convexity, a standard network would only be able to capture monotonically increasing relationships. By including the negative copy of the input, the network gains the ability to learn decreasing physical relationships as well \cite{chen2018optimal}. The network's output $y_t$ is guaranteed to be convex with respect to the input sequence if all associated weight matrices are \textbf{non-negative} and both activation functions ($g_1, g_2$) are \textbf{convex and non-decreasing} \cite{chen2018optimal}.

\subsubsection{Input Convex Long Short-Term Memory}
To combine the powerful temporal modelling capabilities of LSTMs with the advantages of input convexity, Wang et al. proposed the IC-LSTM \cite{wang2025real}. Ensuring the convexity of a standard LSTM is challenging because multiplying its independent gating mechanisms together naturally violates convexity guarantees. The IC-LSTM solves this problem by using \textbf{shared weights} and distinct \textbf{scaling matrices} \cite{wang2025real}. The core computations, including the intermediate dense layer and the final output, are defined as follows:
\begin{align}
    f_t &= g^{(f)}\Bigl( D^{(f)}\bigl(W^{(x)}\hat{x}_t + W^{(h)}h_{t-1}\bigr) + b^{(f)} \Bigr) \label{eq:iclstm_f} \\
    i_t &= g^{(i)}\Bigl( D^{(i)}\bigl(W^{(x)}\hat{x}_t + W^{(h)}h_{t-1}\bigr) + b^{(i)} \Bigr) \label{eq:iclstm_i} \\
    o_t &= g^{(o)}\Bigl( D^{(o)}\bigl(W^{(x)}\hat{x}_t + W^{(h)}h_{t-1}\bigr) + b^{(o)} \Bigr) \label{eq:iclstm_o} \\
    \tilde{c}_t &= g^{(c)}\Bigl( D^{(c)}\bigl(W^{(x)}\hat{x}_t + W^{(h)}h_{t-1}\bigr) + b^{(c)} \Bigr) \label{eq:iclstm_c_tilde} \\
    c_t &= f_t \odot c_{t-1} + i_t \odot \tilde{c}_t \label{eq:iclstm_c} \\
    h_{t,l} &= o_t \odot g^{(h)}(c_t) \label{eq:iclstm_h} \\
    z_{t,l} &= g^{(d)}\bigl(W_l^{(d)}h_{t,l} + b_l^{(d)}\bigr) + \hat{x}_t \label{eq:iclstm_z} \\
    y_t &= g^{(y)}\bigl(W^{(y)}z_{t,L} + b^{(y)}\bigr) \label{eq:iclstm_y}
\end{align}
Here, $f_t, i_t,$ and $o_t$ represent the forget, input, and output gates. The variable $c_t$ is the cell state, $h_{t,l}$ is the hidden state of the $l$-th layer, and $\odot$ denotes element-wise multiplication. Just like the ICRNN, it uses an \textbf{expanded input} $\hat{x}_t = [x_t^\top, -x_t^\top]^\top$. To map the hidden state to the final prediction, the model applies a dense layer $z_{t,l}$ equipped with a parameter-free \textbf{skip connection} that directly adds the expanded input $\hat{x}_t$ to enhance generalization. The final output $y_t$ is then computed from the last layer's representation $z_{t,L}$ \cite{wang2025real}.

Instead of using separate weight matrices for each gate, the IC-LSTM relies on a shared input weight matrix $W^{(x)}$ and a shared hidden weight matrix $W^{(h)}$. To ensure the gates can still learn distinct behaviors, it applies distinct diagonal scaling matrices ($D^{(f)}, D^{(i)}, D^{(o)}, D^{(c)}$). Based on the theoretical proofs by Wang et al. \cite{wang2025real}, an L-layer IC-LSTM model is convex with respect to its input sequence if the following strict conditions hold:
\begin{itemize}
    \item The shared weight matrices ($W^{(x)}, W^{(h)}$), the dense layer weights $W_l^{(d)}$, and the output layer weight $W^{(y)}$, must all be \textbf{non-negative}.
    \item The trainable scaling vectors, represented as diagonal matrices $D^{(\cdot)}$, must be \textbf{non-negative}.
    \item All internal activation functions ($g^{(f)}$, $g^{(i)}$, $g^{(o)}$, $g^{(c)}$, $g^{(h)}$, and $g^{(d)}$) must be \textbf{convex, non-decreasing, AND strictly non-negative}. The final output activation function $g^{(y)}$ must be \textbf{convex and non-decreasing}.
\end{itemize}

The IC-LSTM extends input convexity to a more expressive recurrent architecture and provides an effective model for temporal prediction and optimization. Nevertheless, its hidden and cell states are updated sequentially across the input window, and the length of the unrolled recurrent computation grows with the historical sequence length. Its convexity-compatible gating structure also contains repeated non-negative multiplicative state updates. These properties do not imply that numerical failure must occur, but they motivate a systematic evaluation of computational cost and numerical training behaviour over longer historical sequences. Their practical implications are examined in Sections~\ref{subsubsec:numerical_training_stability} and~\ref{subsubsec:solver_time_reliability}.

\subsubsection{Identifying the Research Gap}
\label{subsubsec:research_gap}

Existing temporal ICNNs demonstrate that neural models can represent sequential dynamical systems while preserving the convexity required for downstream optimization. Feed-forward autoregressive ICNNs and recurrent architectures such as ICRNN and IC-LSTM have therefore established an important basis for learning-based convex MPC~\cite{chen2018optimal,wang2025real}. The more expressive recurrent formulations, however, retain sequential hidden-state propagation across the historical input window.

Transformer encoders provide an alternative non-recurrent mechanism for extracting temporal information and permit the positions within an input sequence to be processed without recurrent state updates~\cite{vaswani2017attention}. Standard Transformer components cannot be transferred directly into an ICNN, because dot-product attention, unconstrained projections, normalisation operations, and conventional feed-forward blocks do not generally preserve input convexity or the monotonicity required for recursive multi-step prediction.

This reveals an architectural gap between temporal input convex modelling and attention-based sequence processing. An input convex encoder architecture must preserve the required convexity and monotonicity properties throughout its input embedding, attention mechanism, residual pathways, feed-forward layers, and output mapping. It must also remain compatible with recursive multi-step substitution when embedded in MPC.

The proposed IC-EoT addresses this gap through a convex additive attention mechanism, non-negative affine mappings, convex non-decreasing activations and residual additions. Its input convex properties are established formally in Section~\ref{subsec:iceot_convexity}. The potential computational and numerical advantages of replacing recurrent state propagation with an encoder-only architecture are not assumed as theoretical consequences of the design; they are evaluated empirically in Section~\ref{subsec:predictive_model_evaluation} and Section~\ref{subsec:mpc_performance_evaluation}.

\section{Input Convex Encoder-only Transformer}
\label{sec:iceot}

This section introduces the proposed IC-EoT, designed to overcome the limitations of recurrent architectures in MPC applications. First, the rationale for adopting an EoT architecture is discussed. Then, the detailed architecture of the IC-EoT is presented, followed by a formal proof of its convex properties. Finally, toy examples are used to visually demonstrate its convexity and evaluate its modelling capability.

\subsection{Architectural Choice: Encoder-Only Transformer}
The original Transformer architecture, as introduced by Vaswani et al. \cite{vaswani2017attention}, comprises both an \textit{encoder} and a \textit{decoder}. This complete structure was primarily developed for sequence-to-sequence tasks, such as machine translation, where an output sequence is generated autoregressively. However, the predictive task within the current MPC framework differs fundamentally from such sequence generation problems. Given a fixed-length historical sequence, the model is required to directly estimate the building states and power consumption at the next time step.

Recent studies demonstrate the effectiveness of Transformer-based architectures for building cooling load and energy consumption forecasting \cite{li2023transformer, spencer2025transfer, gu2025large}. In particular, encoder-based Transformer models have been successfully applied to building energy consumption forecasting and sequential HVAC control \cite{spencer2025transfer, kim2025realtime}. These successful applications strongly support the suitability of self-attention encoders for extracting temporal representations from building operational data.

For the one-step-ahead regression task considered here, the encoder is sufficient to transform the historical input sequence into a context-aware representation. From this representation, the building states and power consumption at the next time step can be directly predicted through a regression head. A separate autoregressive decoder, whose principal role is to generate an output sequence conditioned on previously generated outputs, is therefore not necessary. Accordingly, an EoT architecture is adopted. This design retains the Transformer's ability to capture long-range temporal dependencies while allowing the historical sequence to be processed in parallel. 
As a result, it provides a streamlined architecture for one-step forecasting within MPC.

\subsection{Architecture of the Input Convex Encoder-Only Transformer}
\label{subsec:iceot_architecture}

Recurrent architectures such as RNNs and LSTMs provide an effective means of modelling temporal dependencies, and IC-LSTM extends this capability while preserving input convexity. However, its hidden and cell states must be propagated sequentially across the historical input window, and its recurrent update contains repeated multiplicative gate and state operations. These characteristics motivate the development of a non-recurrent input convex architecture and are examined empirically in this study through numerical training-stability and online MPC-computation experiments.

\begin{figure}[t!]
    \centering
    \includegraphics[width=0.95\columnwidth]{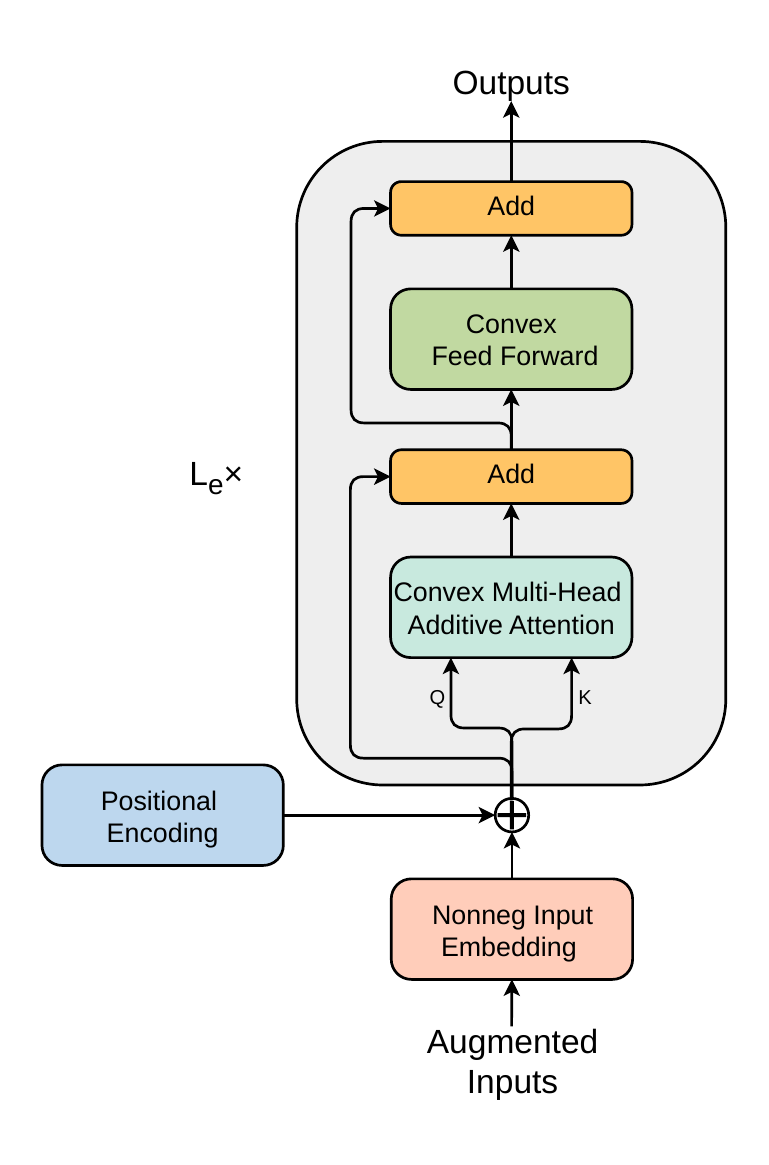}
    \caption{Overall architecture of the proposed IC-EoT. The augmented historical input sequence is first mapped by a non-negative input embedding and combined with positional encoding. The resulting representation is processed by $L_e$ stacked encoder blocks, each consisting of a convex multi-head additive attention layer, residual additions, and a convex feed-forward sub-layer.}
    \label{fig:iceot_encoder_architecture}
\end{figure}

To retain temporal sequence modelling without recurrent state propagation, this work proposes the IC-EoT. As illustrated in Figure~\ref{fig:iceot_encoder_architecture}, the model adopts an encoder-only structure and replaces conventional dot-product self-attention with a convex additive attention mechanism. The attention layer and the surrounding encoder block are constructed from non-negative affine mappings, convex non-decreasing and non-negative activation functions, Hadamard products satisfying the conditions established below, and residual additions.

The data flow begins at the bottom of Figure~\ref{fig:iceot_encoder_architecture}. Let the historical input sequence at time $t$ be
\begin{equation}
X_t=\begin{bmatrix}x_{t-N_h+1}^{\top}\\ \vdots\\ x_t^{\top}\end{bmatrix}\in\mathbb{R}^{N_h\times d_{\mathrm{in}}},
\end{equation}
where $N_h$ denotes the look-back window length and $d_{\mathrm{in}}$ denotes the number of input features. Following the standard construction used in input convex recurrent models, the input is augmented as
\begin{equation}
\widehat X_t=[X_t,-X_t]\in\mathbb{R}^{N_h\times 2d_{\mathrm{in}}}.
\end{equation}
As previously discussed in Section~\ref{Related_work: icrnn}, this augmented representation allows the network to capture decreasing physical relationships. Because all subsequent weight matrices must be non-negative to maintain convexity, including the negative copy of the input ensures that the model retains its full expressive capability without violating these strict structural constraints. As shown in Figure~\ref{fig:iceot_encoder_architecture}, this augmented input is then mapped by a non-negative embedding layer and combined with a positional encoding. The embedded sequence is obtained as
\begin{equation}
H^{(0)}=\widehat X_t W_E+\mathbf{1}b_E^{\top}+P,
\label{eq:iceot_embedding}
\end{equation}
where $W_E\in\mathbb{R}^{2d_{\mathrm{in}}\times d_{\mathrm{model}}}$ is constrained elementwise non-negative, $b_E\in\mathbb{R}^{d_{\mathrm{model}}}$ is an unconstrained bias vector, and $P\in\mathbb{R}^{N_h\times d_{\mathrm{model}}}$ denotes the positional encoding.

\begin{figure*}[t!]
    \centering
    \includegraphics[width=0.95\textwidth]{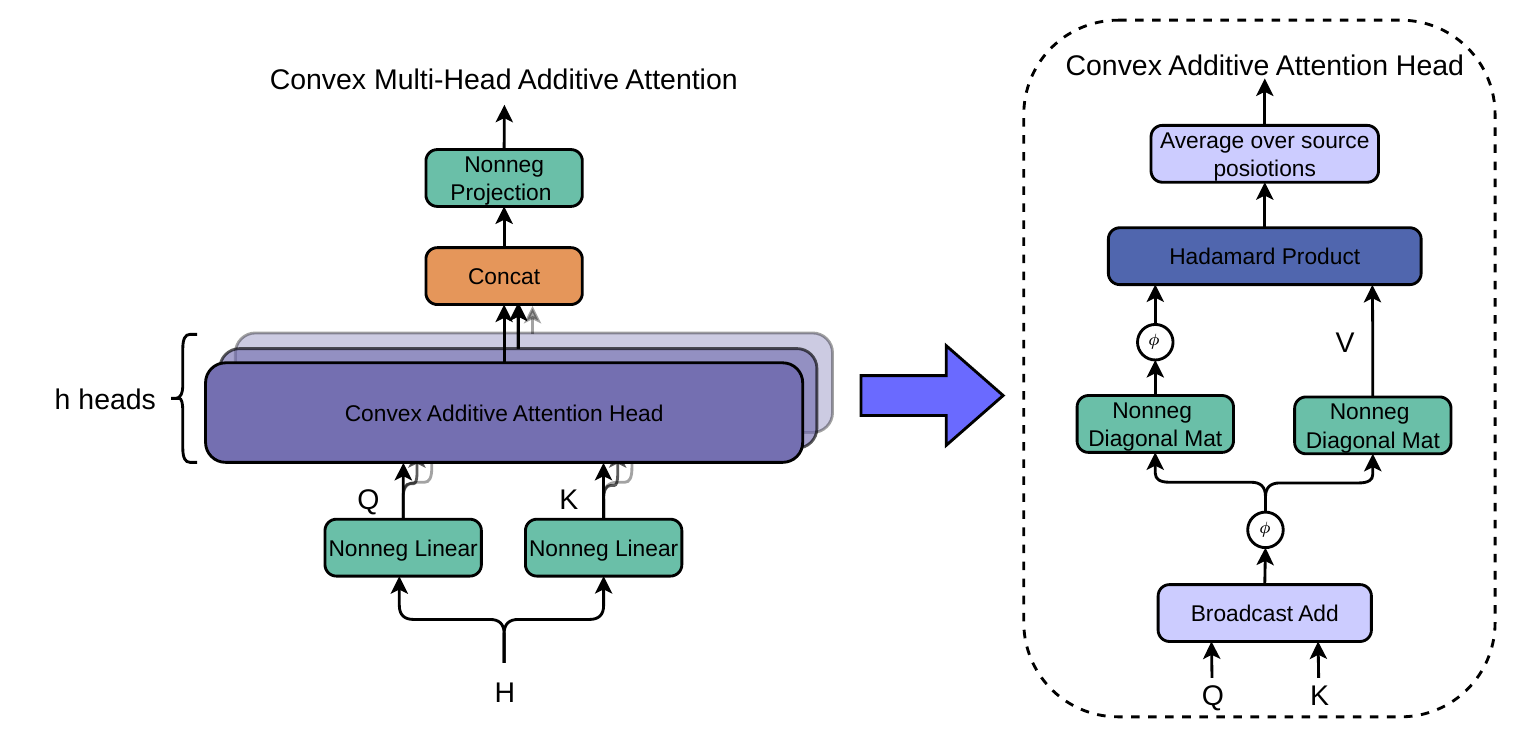}
    \caption{Structure of the proposed convex multi-head additive attention layer. The left panel shows the multi-head wrapper, where $n_{\mathrm{head}}$ convex additive attention heads are concatenated and passed through a non-negative output projection. The right panel zooms in on a single head: non-negative query and key projections are combined through broadcast addition, transformed by convex non-decreasing activations, split into gate and value branches through non-negative diagonal mappings, and finally aggregated by a Hadamard product followed by averaging over source positions.}
    \label{fig:convex_additive_attention}
\end{figure*}

The resulting representation is subsequently processed by a stack of $L_e$ identical encoder blocks. For the $\ell$-th block, where $\ell=1,\ldots,L_e$, and the $p$-th attention head, where $p=1,\ldots,n_{\mathrm{head}}$, let $d_h=d_{\mathrm{model}}/n_{\mathrm{head}}$. The core component of each block is the convex multi-head additive attention layer, detailed in Figure~\ref{fig:convex_additive_attention}. As depicted in the right panel of the figure, the computation within a single attention head begins by generating query and key representations via non-negative linear maps:
\begin{equation}
Q^{(\ell,p)}=H^{(\ell-1)}W_Q^{(\ell,p)},\qquad K^{(\ell,p)}=H^{(\ell-1)}W_K^{(\ell,p)},
\label{eq:iceot_qk}
\end{equation}
where $W_Q^{(\ell,p)},W_K^{(\ell,p)}\in\mathbb{R}^{d_{\mathrm{model}}\times d_h}$ and $W_Q^{(\ell,p)}\ge0$, $W_K^{(\ell,p)}\ge0$. All inequalities are understood elementwise unless otherwise stated.

For query position $i$ and source position $j$, the additive pair representation is defined as
\begin{equation}
S_{ij}^{(\ell,p)}=Q_i^{(\ell,p)}+K_j^{(\ell,p)}+b_Z^{(\ell,p)},
\label{eq:iceot_pair_s}
\end{equation}
\begin{equation}
Z_{ij}^{(\ell,p)}=\phi_Z\left(S_{ij}^{(\ell,p)}\right),
\label{eq:iceot_pair_z}
\end{equation}
where $b_Z^{(\ell,p)}\in\mathbb{R}^{d_h}$ is unconstrained. The activation functions used in the attention construction are chosen to be convex, non-decreasing, and non-negative; ReLU is used in the implementation.

The channel-wise attention gate is defined as
\begin{equation}
A_{ij}^{(\ell,p)}=\phi_A\left(Z_{ij}^{(\ell,p)}D_A^{(\ell,p)}+b_A^{(\ell,p)}\right),
\label{eq:iceot_attention_gate}
\end{equation}
where $D_A^{(\ell,p)}\in\mathbb{R}^{d_h\times d_h}$ is a trainable non-negative diagonal matrix and $b_A^{(\ell,p)}\in\mathbb{R}^{d_h}$ is unconstrained. The value branch is obtained from the same pair representation:
\begin{equation}
V_{ij}^{(\ell,p)}=Z_{ij}^{(\ell,p)}D_V^{(\ell,p)},
\label{eq:iceot_value}
\end{equation}
where $D_V^{(\ell,p)}\in\mathbb{R}^{d_h\times d_h}$ is trainable, diagonal, and elementwise non-negative. The pairwise source contribution is then
\begin{equation}
M_{ij}^{(\ell,p)}=A_{ij}^{(\ell,p)}\odot V_{ij}^{(\ell,p)},
\label{eq:iceot_pair_contribution}
\end{equation}
where $\odot$ denotes the Hadamard product. The context vector of head $p$ at query position $i$ is computed by averaging over all source positions:
\begin{equation}
C_i^{(\ell,p)}=\frac{1}{N_h}\sum_{j=1}^{N_h}M_{ij}^{(\ell,p)}.
\label{eq:iceot_context}
\end{equation}

Returning to the multi-head wrapper shown in the left panel of Figure~\ref{fig:convex_additive_attention}, the individual head outputs are concatenated and passed through a non-negative output projection:
\begin{equation}
O_{\mathrm{att}}^{(\ell)}=\mathrm{Concat}\left(C^{(\ell,1)},\ldots,C^{(\ell,n_{\mathrm{head}})}\right)W_O^{(\ell)}+b_O^{(\ell)},
\label{eq:iceot_att_output}
\end{equation}
where $W_O^{(\ell)}\ge0$ and $b_O^{(\ell)}$ is unconstrained. 

Following the structural flow in Figure~\ref{fig:iceot_encoder_architecture}, a residual connection is applied after the attention layer:
\begin{equation}
\widetilde H^{(\ell)}=H^{(\ell-1)}+O_{\mathrm{att}}^{(\ell)}.
\label{eq:iceot_first_residual}
\end{equation}
The representation is then passed through a position-wise convex feed-forward network:
\begin{equation}
F^{(\ell)}=\phi_F\left(\widetilde H^{(\ell)}W_1^{(\ell)}+b_1^{(\ell)}\right)W_2^{(\ell)}+b_2^{(\ell)},
\label{eq:iceot_ffn}
\end{equation}
where $W_1^{(\ell)}\ge0$ and $W_2^{(\ell)}\ge0$. The block output is obtained by another residual connection:
\begin{equation}
H^{(\ell)}=\widetilde H^{(\ell)}+F^{(\ell)}.
\label{eq:iceot_second_residual}
\end{equation}
Notably, layer normalization is omitted from this architecture. Its input-dependent normalization statistics generally do not preserve input convexity.

After passing through all $L_e$ encoder blocks, the representation corresponding to the most recent time step is selected:
\begin{equation}
h_{\mathrm{last}}=H^{(L_e)}_{(N_h,:)}.
\label{eq:iceot_last_token}
\end{equation}
where the subscript $(N_h,:)$ denotes the $(N_h)-th$ row and all feature dimensions. Finally, the one-step prediction is produced by a non-negative affine readout:
\begin{equation}
\widehat y_{t+1}=h_{\mathrm{last}}W_y+b_y,\qquad W_y\ge0,
\label{eq:iceot_output}
\end{equation}
where $b_y$ is unconstrained. If the network is trained in standardized coordinates, the de-standardization step is a fixed affine transformation with positive scale factors and therefore preserves convexity.

\subsection{Convexity of the Input Convex Encoder-Only Transformer}
\label{subsec:iceot_convexity}

This subsection establishes the componentwise input convexity of the proposed IC-EoT. The proof first isolates the only multiplicative operation in the Convex Multi-Head Additive Attention layer and then propagates the resulting property through the attention layer, encoder blocks, and final readout.

\noindent\textbf{Definition 1 (Componentwise convexity and monotonicity).} A mapping $f:\mathbb{R}^n\rightarrow\mathbb{R}^m$ is componentwise convex if each scalar component $f_r$ is convex. It is componentwise non-decreasing if $u\preceq v\Rightarrow f(u)\preceq f(v)$, where $\preceq$ denotes the elementwise partial order.

\noindent\textbf{Assumption 1 (Structural conditions).} The IC-EoT satisfies the following conditions. First, the matrices $W_E$, $W_Q^{(\ell,p)}$, $W_K^{(\ell,p)}$, $D_A^{(\ell,p)}$, $D_V^{(\ell,p)}$, $W_O^{(\ell)}$, $W_1^{(\ell)}$, $W_2^{(\ell)}$, and $W_y$ are elementwise non-negative for all blocks $\ell$ and heads $p$, and the matrices $D_A^{(\ell,p)}$ and $D_V^{(\ell,p)}$ are diagonal. Second, the activations $\phi_Z$, $\phi_A$, and $\phi_F$ are applied componentwise and are convex, non-decreasing, and non-negative. ReLU is used in the implementation. Third, the positional encoding and all bias vectors are independent of the network input. No sign restriction is imposed on the bias vectors.

\noindent\textbf{Lemma 1 (Closure under non-negative affine maps and monotone convex composition).} Let $f:\mathbb{R}^n\rightarrow\mathbb{R}^m$ be componentwise convex and componentwise non-decreasing. If $A\ge0$, then $Af+b$ is also componentwise convex and componentwise non-decreasing. Moreover, if $\phi$ is componentwise convex and non-decreasing, then $\phi\circ f$ is componentwise convex and componentwise non-decreasing.

\noindent\textit{Proof.} A non-negative weighted sum of convex functions is convex, and a non-negative weighted sum of non-decreasing functions is non-decreasing. The composition result follows from the standard composition rule that a convex non-decreasing outer function composed with a convex inner function remains convex. Monotonicity is preserved because both the outer and inner mappings are componentwise non-decreasing. $\hfill\square$

\noindent\textbf{Lemma 2 (Convexity of the shared latent gate-value product).} Let $z\ge0$, $d_A\ge0$, $d_V\ge0$, and $b_A\in\mathbb{R}$. Let $\phi_A:\mathbb{R}\rightarrow\mathbb{R}_{+}$ be convex, non-decreasing, and non-negative. Define
\[
m(z)=\phi_A(d_Az+b_A)d_Vz.
\label{eq:product_scalar}
\]
Then $m(z)$ is non-negative, non-decreasing, and convex on $\mathbb{R}_{+}$.

\noindent\textit{Proof.} Non-negativity follows directly from $\phi_A(\cdot)\ge0$, $d_V\ge0$, and $z\ge0$. To show monotonicity and convexity, first consider the case where $\phi_A$ is differentiable. Then
\[
m'(z)=d_V\phi_A(d_Az+b_A)+d_Vzd_A\phi_A'(d_Az+b_A)\ge0,
\]
because $d_A,d_V,z\ge0$, $\phi_A\ge0$, and $\phi_A'\ge0$. Similarly,
\[
m''(z)=2d_Vd_A\phi_A'(d_Az+b_A)+d_Vzd_A^2\phi_A''(d_Az+b_A)\ge0,
\]
because $\phi_A$ is convex and non-decreasing. Therefore $m$ is non-decreasing and convex on $\mathbb{R}_{+}$. For non-smooth activations such as ReLU, the same result follows by standard subdifferential arguments or, equivalently, by approximating ReLU with a sequence of smooth convex non-decreasing non-negative functions and using the fact that the pointwise limit of convex non-decreasing functions remains convex and non-decreasing. $\hfill\square$

\noindent\textbf{Proposition 1 (Convexity of the Convex Multi-Head Additive Attention layer).} Suppose $H^{(\ell-1)}$ is componentwise convex and componentwise non-decreasing in the expanded input $\widehat X_t$. Under Assumption 1, the attention output $O_{\mathrm{att}}^{(\ell)}$ defined by Eqs.~\eqref{eq:iceot_qk}--\eqref{eq:iceot_att_output} is also componentwise convex and componentwise non-decreasing in $\widehat X_t$.

\noindent\textit{Proof.} Because $W_Q^{(\ell,p)}\ge0$ and $W_K^{(\ell,p)}\ge0$, Lemma 1 implies that $Q^{(\ell,p)}$ and $K^{(\ell,p)}$ are componentwise convex and componentwise non-decreasing in $\widehat X_t$. Their sum in Eq.~\eqref{eq:iceot_pair_z}, together with the input-independent bias $b_Z^{(\ell,p)}$, preserves these properties. Since $\phi_Z$ is convex, non-decreasing, and non-negative, Lemma 1 further implies that $Z_{ij}^{(\ell,p)}$ is componentwise convex, componentwise non-decreasing, and non-negative. Consider one scalar channel of $Z_{ij}^{(\ell,p)}$, denoted by $z$. The corresponding scalar attention gate and value are $a(z)=\phi_A(d_Az+b_A)$ and $v(z)=d_Vz$, with $d_A,d_V\ge0$. Lemma 2 shows that the scalar product $m(z)=a(z)v(z)$ is non-negative, non-decreasing, and convex in $z$. Since $z$ is itself componentwise convex and componentwise non-decreasing in $\widehat X_t$, Lemma 1 implies that $m(z(\widehat X_t))$ is componentwise convex and componentwise non-decreasing in $\widehat X_t$. Applying this argument componentwise proves that $M_{ij}^{(\ell,p)}=A_{ij}^{(\ell,p)}\odot V_{ij}^{(\ell,p)}$ has the same properties. Equation~\eqref{eq:iceot_context} is an average over source positions and therefore a positive linear combination of the pairwise contributions. Hence each head context $C^{(\ell,p)}$ is componentwise convex and componentwise non-decreasing. Concatenation only reorders scalar components, and the output projection in Eq.~\eqref{eq:iceot_att_output} is non-negative affine. Therefore $O_{\mathrm{att}}^{(\ell)}$ is componentwise convex and componentwise non-decreasing in $\widehat X_t$. $\hfill\square$

\noindent\textbf{Proposition 2 (Input convexity of an IC-EoT encoder block).} If $H^{(\ell-1)}$ is componentwise convex and componentwise non-decreasing in $\widehat X_t$, then the block output $H^{(\ell)}$ defined by Eqs.~\eqref{eq:iceot_first_residual}--\eqref{eq:iceot_second_residual} has the same properties.

\noindent\textit{Proof.} By Proposition 1, $O_{\mathrm{att}}^{(\ell)}$ is componentwise convex and componentwise non-decreasing in $\widehat X_t$. The residual addition in Eq.~\eqref{eq:iceot_first_residual} preserves both properties. The feed-forward sub-layer consists of a non-negative affine map, a componentwise convex non-decreasing activation, and another non-negative affine map. By repeated application of Lemma 1, $F^{(\ell)}$ is componentwise convex and componentwise non-decreasing in $\widehat X_t$. The second residual addition in Eq.~\eqref{eq:iceot_second_residual} again preserves these properties. Therefore $H^{(\ell)}$ is componentwise convex and componentwise non-decreasing in $\widehat X_t$. $\hfill\square$

\noindent\textbf{Theorem 1 (Input convexity of the complete IC-EoT).} Under Assumption 1, an IC-EoT with $L_e$ encoder blocks defines a componentwise convex and componentwise non-decreasing mapping from the expanded input $\widehat X_t$ to the normalized one-step prediction $\widehat y^{\,s}_{t+1}$.

\noindent\textit{Proof.} The embedding in Eq.~\eqref{eq:iceot_embedding} is a non-negative affine map of $\widehat X_t$ plus an input-independent positional encoding. Hence $H^{(0)}$ is componentwise convex and componentwise non-decreasing in $\widehat X_t$. Applying Proposition 2 inductively for $\ell=1,\ldots,L_e$ shows that $H^{(L_e)}$ has the same properties. Selecting the last token in Eq.~\eqref{eq:iceot_last_token} is a coordinate projection and therefore preserves them. Finally, Eq.~\eqref{eq:iceot_output} is a non-negative affine map. Thus every component of $\widehat y^{\,s}_{t+1}$ is convex and non-decreasing in $\widehat X_t$. $\hfill\square$

\noindent\textbf{Corollary 1 (Convexity with respect to the original input sequence).} The IC-EoT prediction is componentwise convex with respect to the original sequence $X_t$.

\noindent\textit{Proof.} The expanded input $\widehat X_t=[X_t,-X_t]$ is an affine transformation of $X_t$. The composition of a convex function with an affine mapping is convex. Therefore, Theorem 1 implies componentwise convexity with respect to the original sequence $X_t$. The expanded representation allows both increasing and decreasing dependencies with respect to the original variables; hence monotonicity is guaranteed with respect to $\widehat X_t$, but is not claimed with respect to $X_t$. $\hfill\square$

\subsection{Toy Examples: Surface Fitting} \label{sec:toy_example}
To visually demonstrate the input convexity of the proposed IC-EoT, a 2-dimensional regression task is conducted. To facilitate a direct and fair comparison with prior work, this experiment uses the same three challenging, non-convex bivariate scalar functions as those in Wang et al.~\cite{wang2025real}:
\begin{align}
    f_1(x,y) &= -\cos(4x^2 + 4y^2) \\
    f_2(x,y) &= \max(\min(x^2+y^2, (2x-1)^2+(2y-1)^2-2), \nonumber \\ 
             &\quad -(2x+1)^2-(2y+1)^2+4) \\
    f_3(x,y) &= x^2(4-2.1x^2+x^{4/3}) - 4y^2(1-y^2) + xy
\end{align}

\begin{figure}[t!]
    \centering
    \includegraphics[width=0.95\columnwidth]{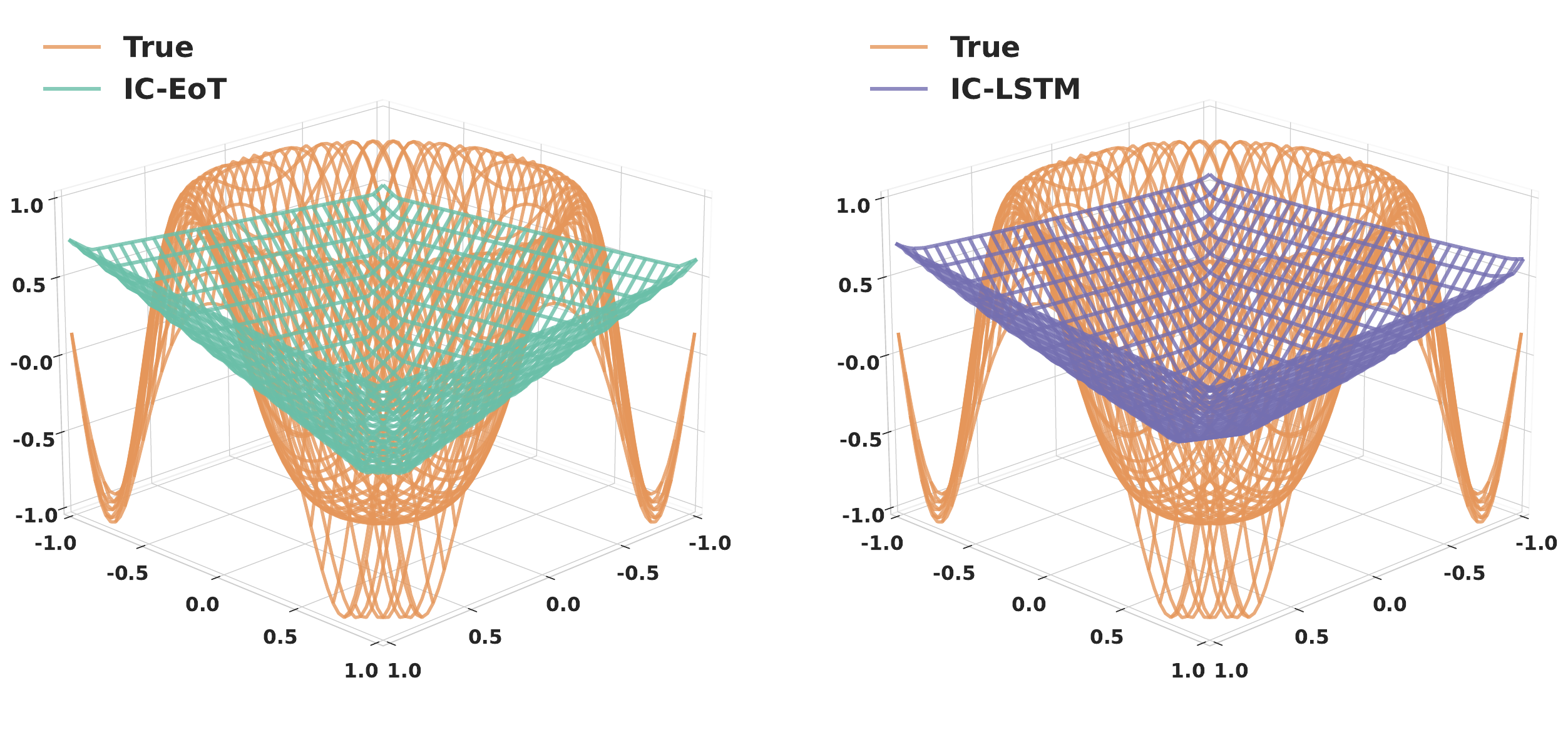}
    \caption{Fitting result for function $f_1(x,y)$. The left panel shows the comparison between the true surface and the IC-EoT prediction. The right panel shows the same comparison for IC-LSTM.}
    \label{fig:f1_comparison}
\end{figure}

\begin{figure}[t!]
    \centering
    \includegraphics[width=0.95\columnwidth]{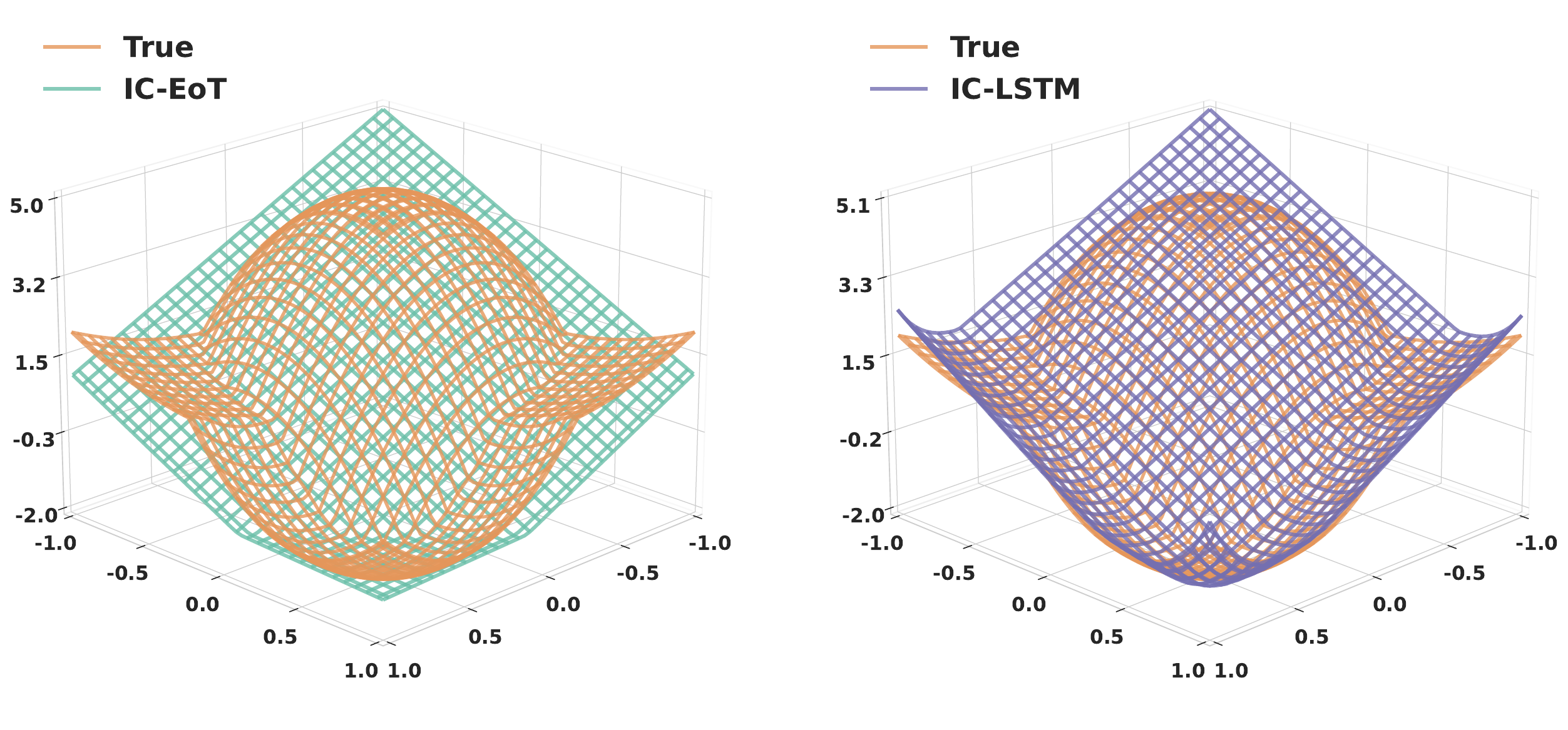}
    \caption{Fitting result for function $f_2(x,y)$.}
    \label{fig:f2_comparison}
\end{figure}

\begin{figure}[t!]
    \centering
    \includegraphics[width=0.95\columnwidth]{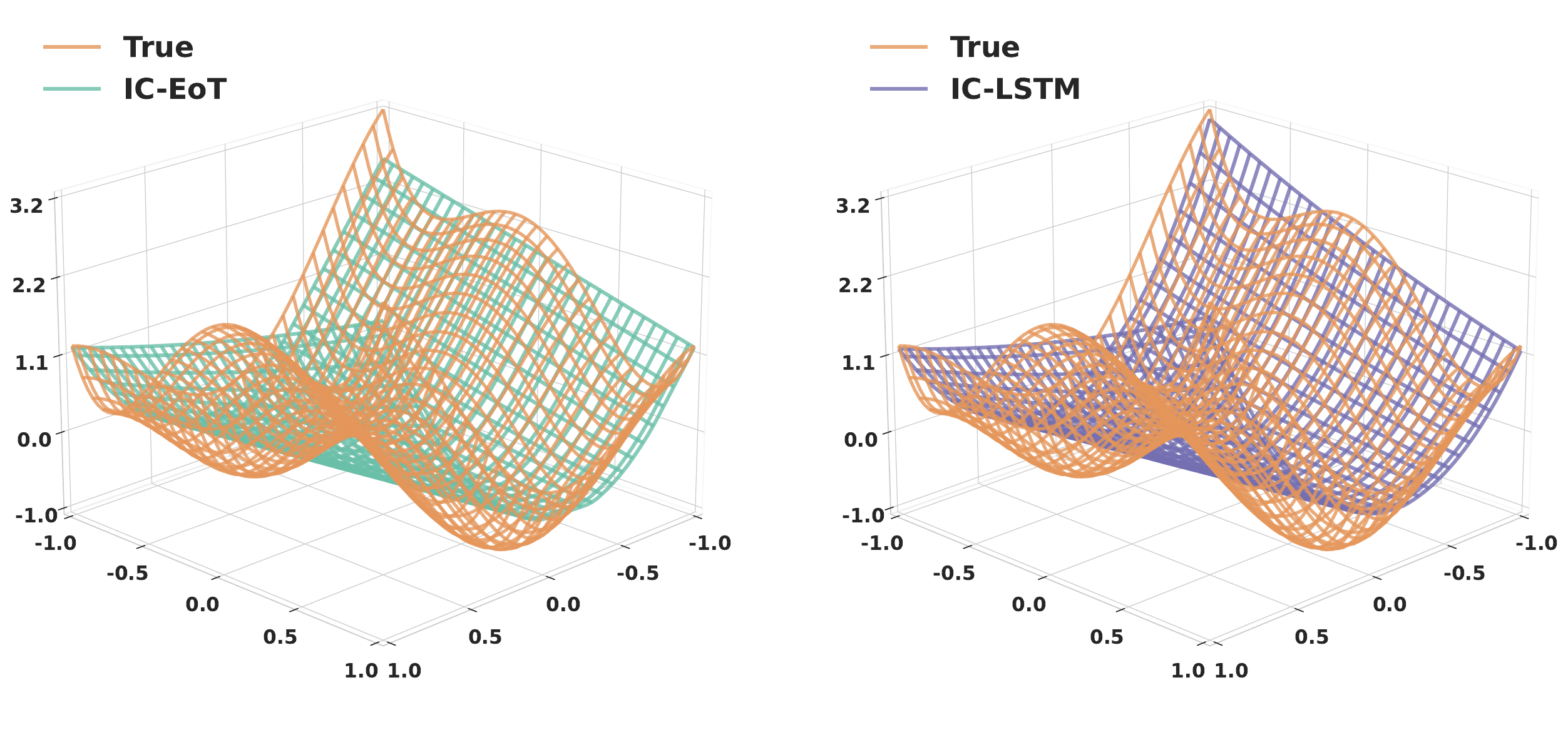}
    \caption{Fitting result for function $f_3(x,y)$.}
    \label{fig:f3_comparison}
\end{figure}

Given their input convex architectures, both IC-EoT and IC-LSTM are expected to learn convex approximations of these inherently non-convex surfaces. The results are presented in Figures~\ref{fig:f1_comparison}, \ref{fig:f2_comparison}, and \ref{fig:f3_comparison}. As shown, both models successfully produce convex surfaces as predicted by their theoretical properties. For the highly oscillatory cosine function $f_1$, the models capture the global bowl shape but are unable to fit the local non-convex ripples.

The quantitative performance metrics are summarized in Table~\ref{tab:toy_results}. The results indicate that the proposed IC-EoT achieves a modelling accuracy comparable to that of IC-LSTM. Specifically, the IC-EoT performs slightly better on functions $f_1$ and $f_3$, while the IC-LSTM shows a slight advantage on function $f_2$. This is a crucial finding, as it demonstrates that the architectural shift from a recurrent to a parallel, attention-based structure does not compromise the model's fundamental ability to approximate complex functions within the constraints of convexity.

This toy experiment is intended as a preliminary illustration of the input convex behavior and approximation capability of the IC-EoT, rather than as a comprehensive computational benchmark. A systematic comparison of the two architectures in terms of predictive accuracy, training behavior, computational efficiency, and control performance within the MPC framework is presented in the case study in Section~\ref{sec:case_study}. 

\begin{table}[h!]
\centering
\caption{Performance Comparison on Toy Example Surface Fitting}
\label{tab:toy_results}
\resizebox{0.9\columnwidth}{!}{%
\begin{tabular}{@{}llcc@{}}
\toprule
\textbf{Function} & \textbf{Model} & \textbf{Test MSE} & \textbf{R² Score} \\ \midrule
\multirow{2}{*}{$f_1$ (Cosine)} & IC-EoT & \textbf{0.3924} & \textbf{0.2345} \\
 & IC-LSTM & 0.3975 & 0.2245 \\ \midrule
\multirow{2}{*}{$f_2$ (Complex)} & IC-EoT & 0.5810 & 0.8051 \\
 & IC-LSTM & \textbf{0.4915} & \textbf{0.8352} \\ \midrule
\multirow{2}{*}{$f_3$ (Six-hump)} & IC-EoT & \textbf{0.1490} & \textbf{0.8154} \\
 & IC-LSTM & 0.1552 & 0.8077 \\ \bottomrule
\end{tabular}%
}
\end{table}

\section{Case Study}
\label{sec:case_study}

This section evaluates the proposed IC-EoT in two EnergyPlus-based building testbeds implemented through the Energym co-simulation interface: the residential ApartmentsThermal environment and the commercial OfficesThermostat environment. IC-EoT is compared with the recurrent IC-LSTM and with the conventional EoT and LSTM in terms of predictive accuracy, training behaviour, online MPC solution time, solver termination, and closed-loop cost and comfort performance.

The case study first formulates the soft-constrained input convex MPC problem and then introduces the two building testbeds and their co-simulation framework. The predictive models are subsequently evaluated using common architecture-paired configurations. Multi-seed experiments assess the numerical training stability of IC-EoT and IC-LSTM over a range of historical sequence lengths, after which all four models are embedded in MPC and evaluated over prediction horizons from one to eight hours.

Across the two testbeds, IC-EoT achieves predictive accuracy close to that of IC-LSTM and records lower mean per-epoch training times under the tested configuration. No non-finite training event is observed in the 60 IC-EoT multi-seed runs, whereas IC-LSTM exhibits repeated numerical anomalies at longer historical sequence lengths. At the eight-hour MPC horizon, IC-EoT requires approximately $5.8\times$ and $5.3\times$ less solution time than IC-LSTM in ApartmentsThermal and OfficesThermostat, respectively. It also achieves valid-termination rates of 99.9\% and 100.0\% while providing closed-loop cost and comfort performance comparable to the recurrent input convex baseline. Because ApartmentsThermal does not provide active mechanical cooling, its closed-loop results are interpreted as a reference under limited cooling authority, while OfficesThermostat serves as the principal active-cooling control experiment.

\subsection{MPC Optimization Problem}
\label{subsec:mpc_optimization_problem}

At each control interval, the learning-based MPC controller determines a sequence of future thermostat setpoints over a finite prediction horizon. The neural predictor is denoted generically by $f_{\theta}$ and is instantiated as IC-EoT, IC-LSTM, EoT, or LSTM in the comparative experiments. The convexity result developed in this subsection applies when $f_{\theta}$ is an input convex and componentwise non-decreasing predictor, such as the proposed IC-EoT or the IC-LSTM. The same optimization structure is also used with the conventional EoT and LSTM models, although no convexity guarantee is available for these two baselines.

Let $N_p$ denote the prediction horizon and define the stage-index set
\begin{equation}
\mathcal{K}:=\{0,1,\ldots,N_p-1\}.
\end{equation}
The sequence of future control actions is
\begin{equation}
U_t:=\left\{
u_t,u_{t+1},\ldots,u_{t+N_p-1}
\right\},
\end{equation}
where $u_{t+k}\in\mathbb{R}^{n_u}$ contains the thermostat setpoints applied to the controlled thermal zones or zone groups.

For the input convex predictors, only the control variables are augmented by their negative counterparts:
\begin{equation}
\widetilde u_{t+k}
:=
\begin{bmatrix}
u_{t+k}\\
-u_{t+k}
\end{bmatrix}
=
A_u u_{t+k},
\qquad
A_u:=
\begin{bmatrix}
I_{n_u}\\
-I_{n_u}
\end{bmatrix}.
\label{eq:selective_control_extension}
\end{equation}

For physical interpretation, the recursively propagated model output is partitioned as
\[
\widehat y_{k|t}
=
\begin{bmatrix}
\widehat x_{k|t}\\
\widehat d_{k|t}
\end{bmatrix},
\]
where $\widehat x_{k|t}$ contains the endogenous building variables and $\widehat d_{k|t}$ contains variables that are exogenous from a physical perspective. Both components are generated recursively by the neural predictor and are therefore functions of the preceding control decisions; the partition is introduced only to distinguish their physical roles.

Accordingly, one row of the model input sequence takes the generic form
\begin{equation}
\xi_{k|t}
:=
\begin{bmatrix}
\widehat y_{k|t}\\
\widetilde u_{t+k}
\end{bmatrix},
\qquad
\widehat y_{0|t}=y_t,
\label{eq:selectively_extended_input}
\end{equation}
where $y_t$ denotes the measured model-variable vector at the current control instant. 

The construction in Eq.~\eqref{eq:selective_control_extension} follows the control-oriented input extension introduced by Chen et al.~\cite{chen2018optimal}. It enables the network to represent both increasing and decreasing dependencies on the original control variables while retaining non-negative network weights. Importantly, the recursively propagated model variables $\widehat y_{k|t}$ are not duplicated. Since $\widehat y_{k|t}$ is generated recursively by the predictor, it is generally a componentwise convex function of the preceding control actions, whereas $-\widehat y_{k|t}$ is generally componentwise concave. Including both quantities in subsequent model inputs would therefore invalidate the monotone convex-composition argument required for multi-step prediction. In contrast, both $u_{t+k}$ and $-u_{t+k}$ are affine functions of the original decision variable and can be included without affecting convexity. The unextended recursive pathway therefore preserves the componentwise nondecreasing dependence required when predicted variables are substituted into subsequent prediction steps.

Let $\mathcal{H}_{0|t}$ denote the measured historical sequence available at time $t$, which is fixed when the MPC problem is formulated. Subsequent model-input windows are generated recursively as
\begin{align}
\mathcal{H}_{k+1|t}
&=
\operatorname{shift}
\left(
\mathcal{H}_{k|t},
\xi_{k|t}
\right),
\label{eq:mpc_history_update}\\
\widehat y_{k+1|t}
&=
f_{\theta}
\left(
\mathcal{H}_{k+1|t}
\right),
\qquad k\in\mathcal{K},
\label{eq:mpc_recursive_prediction}
\end{align}
where $\operatorname{shift}(\cdot)$ removes the oldest row and appends the newly constructed row $\xi_{k|t}$. The vector $\widehat y_{k+1|t}$ contains all one-step-ahead variables propagated by the predictive model, including the zone temperatures, electricity consumption, and any additional variables required to construct the subsequent model-input window. The corresponding components are denoted by
\begin{equation}
\widehat T_{z,k+1|t},
\qquad
z\in\mathcal{Z}:=\{1,\ldots,n_z\},
\end{equation}
and
\begin{equation}
\widehat E_{k+1|t},
\end{equation}
respectively.

After recursively substituting Eqs.~\eqref{eq:mpc_history_update}--\eqref{eq:mpc_recursive_prediction}, the predicted quantities can be written directly as functions of the control sequence:
\begin{equation}
\widehat T_{z,k+1|t}
=
\widehat T_{z,k+1|t}(U_t),
\qquad
\widehat E_{k+1|t}
=
\widehat E_{k+1|t}(U_t).
\label{eq:reduced_mpc_predictions}
\end{equation}
This reduced representation avoids treating the nonlinear neural-network dynamics as equality constraints in the optimization problem.

The closed-loop experiments are conducted under summer operating conditions, for which overheating constitutes the relevant thermal-comfort violation. Accordingly, only an upper comfort limit is imposed. An explicit lower temperature limit is omitted because unnecessary cooling increases the predicted electricity cost under the non-negative electricity prices considered in this study, while the lower bounds imposed on the thermostat setpoints further restrict excessive cooling. These mechanisms discourage overcooling, although they should not be interpreted as providing a hard lower-temperature guarantee.

Non-negative slack variables are introduced to soften the upper comfort limit and improve numerical feasibility in the presence of prediction errors or temporarily unavoidable overheating. This formulation follows the soft-constrained ICNN-based building MPC framework in~\cite{bunning2021input}. Since the predicted temperature is convex in the control sequence and the slack variable enters the constraint affinely, the softened upper-temperature constraint remains convex.

Define the sequence of zone-wise slack variables as
\begin{equation}
\mathcal{E}_t
:=
\left\{
\epsilon_{1|t},
\epsilon_{2|t},
\ldots,
\epsilon_{N_p|t}
\right\},
\qquad
\epsilon_{k+1|t}\in\mathbb{R}^{n_z}_{+}.
\end{equation}
The resulting soft-constrained MPC problem is
\begin{equation}
\begin{aligned}
\underset{U_t,\mathcal{E}_t}{\operatorname{minimize}}
\quad
& \sum_{k\in\mathcal{K}}
\left[
\pi_{t+k}\widehat E_{k+1|t}(U_t)
+
\lambda_{\epsilon}
\left\|
\epsilon_{k+1|t}
\right\|_2^2
\right]
\\[1mm]
\operatorname{subject\ to}
\quad
& u_{\min}
\preceq
u_{t+k}
\preceq
u_{\max},
&& k\in\mathcal{K},
\\
& \widehat T_{z,k+1|t}(U_t)
-
\epsilon_{z,k+1|t}
\leq
T_{z,\max,t+k+1},
&& z\in\mathcal{Z},\; k\in\mathcal{K},
\\
& 0
\leq
\epsilon_{z,k+1|t}
\leq
\epsilon_{\max},
&& z\in\mathcal{Z},\; k\in\mathcal{K}.
\end{aligned}
\label{eq:soft_constrained_mpc}
\end{equation}
Here, $\pi_{t+k}\geq0$ is the known electricity price over the corresponding control interval, $u_{\min}$ and $u_{\max}$ are the admissible thermostat-setpoint bounds, $T_{z,\max,t+k+1}$ is the upper comfort limit of zone $z$, and $\lambda_{\epsilon}\geq0$ is the slack-penalty coefficient. The time-of-use tariffs used to determine $\pi_{t+k}$ are specified in Section~\ref{subsubsec:mpc_scenarios_tariffs}. The finite bound $\epsilon_{\max}$ is included for numerical conditioning and to prevent unrealistically large comfort relaxation.

\medskip
\noindent\textbf{Corollary 2 (Convexity of the soft-constrained ICNN--MPC problem).}
Suppose that every component of the neural predictor $f_{\theta}$ is convex in its input sequence. Assume also that $f_{\theta}$ is componentwise non-decreasing with respect to the recursively substituted predicted variables. If the measured initial history $\mathcal{H}_{0|t}$ is fixed, $\pi_{t+k}\geq0$, and $\lambda_{\epsilon}\geq0$, then, under the selective control extension in Eq.~\eqref{eq:selective_control_extension}, the MPC problem in Eq.~\eqref{eq:soft_constrained_mpc} is convex in the joint decision variables $(U_t,\mathcal{E}_t)$.

\medskip
\noindent\textit{Proof.}
At the first prediction step, the measured historical sequence $\mathcal{H}_{0|t}$ and the current model-variable vector $\widehat y_{0|t}=y_t$ are fixed. The only decision-dependent component of the newly appended input row is
\begin{equation}
\widetilde u_t=A_u u_t,
\end{equation}
which is affine in the control decision. Consequently, $\mathcal{H}_{1|t}$ is an affine function of $u_t$. Since every component of $f_{\theta}$ is convex in its input sequence, every component of $\widehat y_{1|t}$ is convex in $u_t$.

Assume that the recursively generated predictions up to stage $k$ are componentwise convex functions of the control sequence $U_t$. The newly appended history row contains the recursively predicted vector $\widehat y_{k|t}$ and the affine extended control $\widetilde u_{t+k}$. The history-shift operation performs only coordinate selection and concatenation and therefore preserves componentwise convexity. Since $f_{\theta}$ is componentwise convex in its input sequence and componentwise non-decreasing with respect to the recursively substituted predicted variables, the monotone convex-composition rule implies that $\widehat y_{k+1|t}$ is also componentwise convex in $U_t$. By induction, every predicted zone temperature $\widehat T_{z,k+1|t}(U_t)$ and electricity-consumption term $\widehat E_{k+1|t}(U_t)$ is convex in the complete control sequence.

Since $\pi_{t+k}\geq0$, the electricity-cost term is a non-negative weighted sum of convex functions. The quadratic slack penalty is convex for $\lambda_{\epsilon}\geq0$. The control and slack bounds are affine constraints. Finally, each softened upper-temperature constraint can be written as
\begin{equation}
\widehat T_{z,k+1|t}(U_t)
-
\epsilon_{z,k+1|t}
-
T_{z,\max,t+k+1}
\leq0.
\end{equation}
Its left-hand side is convex in $(U_t,\mathcal{E}_t)$ because it is the sum of a convex predicted-temperature function and an affine function of the slack variable. Hence, it defines a convex sublevel set. The objective and all feasible-set constraints in Eq.~\eqref{eq:soft_constrained_mpc} are therefore convex, which proves the result.
\hfill$\square$

For IC-EoT- and IC-LSTM-based MPC, Corollary~2 establishes that any local optimum of the reduced problem is also globally optimal. When the conventional EoT or LSTM is used as $f_{\theta}$, the same objective and constraints are retained for a consistent empirical comparison, but the predicted quantities are generally non-convex functions of $U_t$ and the global convexity guarantee no longer applies.

\subsection{Building Testbeds and Co-Simulation Framework}
\label{subsec:testbeds_framework}
A unified co-simulation framework is used for both the offline development of the neural predictive models and their subsequent closed-loop evaluation within MPC. The building dynamics are simulated using \textbf{EnergyPlus}~\cite{crawley2001energyplus}, a widely adopted whole-building energy simulation program. The interaction between EnergyPlus and the Python-based experimental pipeline is implemented using \textbf{Energym}~\cite{scharnhorst2021energym}. Energym is an open-source building simulation library developed specifically for systematic and reproducible controller benchmarking. It provides a common interface through which a Python program can transmit control actions to a building model and receive the resulting building states, energy measurements, and weather information.

Two Energym environments are considered: the residential \texttt{ApartmentsThermal-v0} environment and the commercial \texttt{OfficesThermostat-v0} environment. These environments are used both to generate the simulation trajectories required for predictive-model development and to evaluate the trained models in closed-loop MPC.

\begin{figure*}[t]
    \centering
    \includegraphics[width=\textwidth]{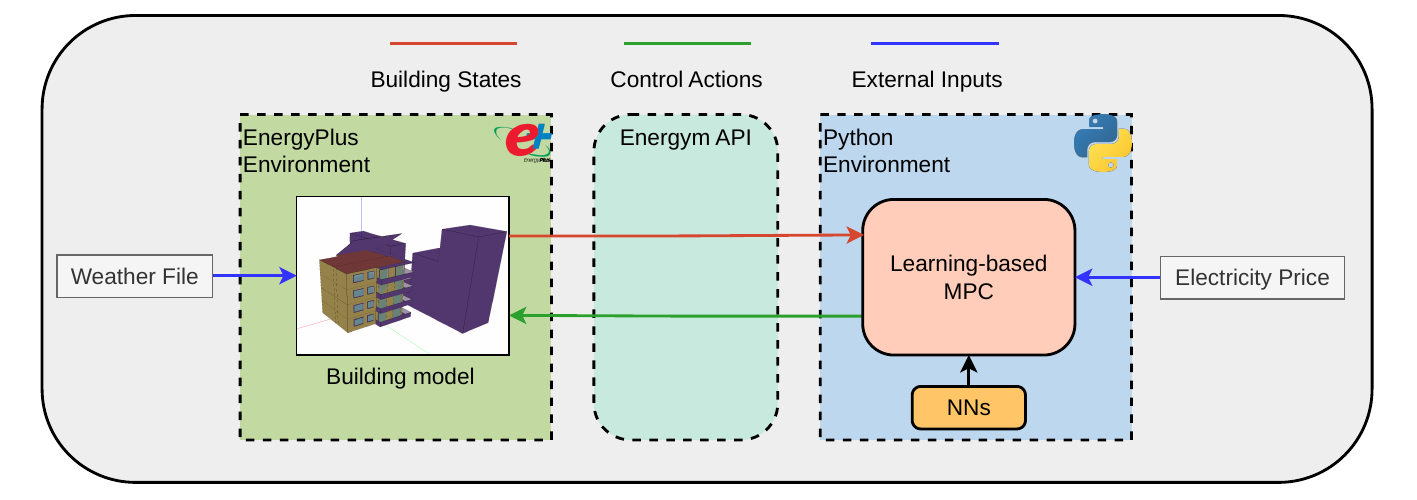}
    \caption{Closed-loop co-simulation framework used for MPC evaluation. The generic EnergyPlus building model is instantiated as either the ApartmentsThermal or OfficesThermostat testbed and communicates with the Python environment through the Energym interface. At each control instant, building observations are returned to the learning-based MPC controller, while the computed thermostat setpoints are applied to the building model. Weather conditions and electricity prices are supplied as external inputs to the simulation and controller, respectively.}
    \label{fig:simulation_framework}
\end{figure*}

Figure~\ref{fig:simulation_framework} illustrates the closed-loop co-simulation architecture. The generic \textit{Building Model} block is instantiated as either ApartmentsThermal or OfficesThermostat. At each control instant, the EnergyPlus model returns the current building observations to the Python environment through the Energym interface. The learning-based MPC controller then uses the selected neural predictor, IC-EoT, IC-LSTM, EoT, or LSTM, to determine the future thermostat setpoint sequence and applies the first control action to the building model. EnergyPlus then advances the building simulation and returns the updated observations, thereby completing the receding-horizon control loop.

The \texttt{ApartmentsThermal-v0} environment represents a four-storey residential building located in Tarragona, Spain, with a total floor area of approximately $417.12~\mathrm{m^2}$ and an air volume of $1042.83~\mathrm{m^3}$~\cite{scharnhorst2021energym}. Each storey corresponds to an individual apartment and contains two thermal zones, resulting in eight zones in total. A single thermostat setpoint is assigned to each apartment, giving four independently controllable thermostat setpoints.

Space heating and domestic hot-water production are supplied by a centralised water-to-water geothermal heat pump connected to a vertical ground heat exchanger. Indoor conditioning is provided through fan-coil units, with two units installed in each apartment. The model does not contain an active mechanical cooling system. Instead, its original summer operation relies on a free-cooling representation implemented through increased air infiltration. 
Consequently, thermostat adjustments provide only limited cooling authority during the summer experiments considered in this study.

The \texttt{OfficesThermostat-v0} environment represents an office building located in Greece, with a total floor area of approximately $643.73~\mathrm{m^2}$ and an air volume of $2504.54~\mathrm{m^3}$~\cite{scharnhorst2021energym}. The building contains 25 conditioned spaces, including office rooms, seminar rooms, a meeting room, circulation areas, storage rooms, bathrooms, a kitchen, and an HVAC equipment room. Among these spaces, 14 thermal zones are equipped with independently controllable thermostat setpoints and are therefore included as controlled zones in this study.

The conditioned spaces are served by water-to-air fan-coil units connected to a two-pipe water distribution system. Depending on the operating mode, the circuit is supplied either with hot water from an oil boiler or with chilled water from an electric air-to-water chiller; simultaneous heating and cooling are not permitted. For the summer experiments considered in this work, the building is operated in cooling mode, with the chiller available and the boiler disabled. The Offices environment therefore provides a higher-dimensional commercial testbed with active mechanical cooling and a larger number of independently controlled thermal zones.

Evaluating both a residential and a commercial testbed allows the predictive and computational findings to be examined across two distinct building types, HVAC configurations, and control dimensions, rather than being based on a single simulation environment.

For both testbeds, the predictive-model datasets are constructed at a 15-minute temporal resolution using trajectories generated from the corresponding EnergyPlus environments. Different weather files are used for offline model development and closed-loop MPC evaluation. The model inputs, prediction targets, and input convex extensions adopted for the two testbeds are detailed in Section~\ref{subsubsec:data_representation}.

\subsection{Predictive Model Configuration, Performance, and Numerical Stability}
\label{subsec:predictive_model_evaluation}

This subsection evaluates the predictive models used in the learning-based MPC framework. The input and output representations adopted for the two building testbeds are first introduced, followed by the model configurations and the common training protocol. The four candidate architectures are then compared in terms of predictive accuracy and per-epoch training time. Finally, multi-seed experiments are conducted to evaluate the numerical training stability of the two input convex architectures across a range of historical sequence lengths.

\subsubsection{Data Representation and Model Inputs}
\label{subsubsec:data_representation}

For both testbeds, the predictive models receive a fixed-length historical sequence and estimate the building variables at the next prediction step. The modelled variables and control inputs used in each testbed are defined below.

For the ApartmentsThermal testbed, the modelled variable vector is
\begin{equation}
\boldsymbol{y}^{\mathrm{apt}}_t =
\begin{bmatrix}
T_{1,t},\ldots,T_{8,t},
E^{\mathrm{fac}}_t,
T^{\mathrm{HP,ret}}_t,
E^{\mathrm{appl}}_t
\end{bmatrix}^{\top},
\label{eq:apartment_model_variables}
\end{equation}
where $T_{1,t},\ldots,T_{8,t}$ are the eight zone temperatures, $E^{\mathrm{fac}}_t$ is the facility electricity consumption, $T^{\mathrm{HP,ret}}_t$ is the heat-pump return-water temperature, and $E^{\mathrm{appl}}_t$ is the appliance electricity consumption. The control vector contains four apartment-level thermostat setpoints:
\begin{equation}
\boldsymbol{u}^{\mathrm{apt}}_t =
\begin{bmatrix}
u_{1,t},u_{2,t},u_{3,t},u_{4,t}
\end{bmatrix}^{\top}.
\label{eq:apartment_control_variables}
\end{equation}

For the OfficesThermostat testbed, the modelled variable vector is
\begin{equation}
\boldsymbol{y}^{\mathrm{off}}_t =
\begin{bmatrix}
T_{1,t},\ldots,T_{14,t},
E^{\mathrm{fac}}_t,
T^{\mathrm{ext}}_t
\end{bmatrix}^{\top},
\label{eq:office_model_variables}
\end{equation}
where the first 14 components are the temperatures of the independently controlled zones, $E^{\mathrm{fac}}_t$ is the facility electricity consumption, and $T^{\mathrm{ext}}_t$ is the external air temperature. The control vector contains 14 zone-level thermostat setpoints:
\begin{equation}
\boldsymbol{u}^{\mathrm{off}}_t =
\begin{bmatrix}
u_{1,t},\ldots,u_{14,t}
\end{bmatrix}^{\top}.
\label{eq:office_control_variables}
\end{equation}

The conventional EoT and LSTM models use the input row
\begin{equation}
\boldsymbol{\zeta}_t =
\begin{bmatrix}
\boldsymbol{y}^{\top}_t &
\boldsymbol{u}^{\top}_t
\end{bmatrix}^{\top},
\label{eq:standard_model_input}
\end{equation}
whereas IC-EoT and IC-LSTM use the control-augmented input row
\begin{equation}
\widetilde{\boldsymbol{\zeta}}_t =
\begin{bmatrix}
\boldsymbol{y}^{\top}_t &
\boldsymbol{u}^{\top}_t &
-\boldsymbol{u}^{\top}_t
\end{bmatrix}^{\top}.
\label{eq:icnn_model_input}
\end{equation}

Only the control variables are duplicated; the measured and recursively predicted building variables are not negatively duplicated. This representation is consistent with the convex MPC formulation in Section~\ref{subsec:mpc_optimization_problem}. The resulting input and output dimensions are summarised in Table~\ref{tab:model_input_output}.

\begin{table*}[t]
    \centering
    \caption{Predictive-model inputs and outputs for the two building testbeds. The ICNN input dimension includes the negative copy of the control variables only.}
    \label{tab:model_input_output}
    \small
    \setlength{\tabcolsep}{4pt}
    \renewcommand{\arraystretch}{1.15}
    \resizebox{\textwidth}{!}{%
    \begin{tabular}{l p{6.7cm} p{4.9cm} c c c}
        \toprule
        \textbf{Testbed}
        & \textbf{Modelled variables and prediction targets}
        & \textbf{Control variables}
        & \textbf{Standard input}
        & \textbf{ICNN input}
        & \textbf{Output} \\
        \midrule
        ApartmentsThermal
        & Eight zone temperatures, facility electricity consumption, heat-pump return-water temperature, and appliance electricity consumption
        & Four apartment-level thermostat setpoints
        & 15 & 19 & 11 \\
        \addlinespace
        OfficesThermostat
        & Fourteen zone temperatures, facility electricity consumption, and external air temperature
        & Fourteen zone-level thermostat setpoints
        & 30 & 44 & 16 \\
        \bottomrule
    \end{tabular}%
    }
\end{table*}

\subsubsection{Model Configurations and Training Protocol}
\label{subsubsec:model_configuration}

Four architectures are considered: the proposed IC-EoT, IC-LSTM, conventional EoT, and conventional LSTM. The EoT and LSTM are used as non-convex architectural baselines for IC-EoT and IC-LSTM, respectively. To provide structurally meaningful comparisons, IC-EoT and EoT use the same encoder depth, attention-head count, model dimension, and feed-forward dimension, while IC-LSTM and LSTM use the same recurrent depth and hidden-state dimension.

The resulting model configurations and trainable parameter counts are summarised in Table~\ref{tab:model_hyperparameters}. The two input convex models have similar parameter counts in both testbeds. The parameter counts of the paired convex and conventional models are not necessarily identical because the convexity constraints alter the admissible weight structures and parameter-sharing mechanisms. In particular, the conventional LSTM contains more parameters than IC-LSTM because the standard LSTM uses separate weight matrices for its gates, whereas the adopted IC-LSTM uses shared non-negative matrices together with gate-specific diagonal scaling vectors.

\begin{table*}[t]
    \centering
    \caption{Model configurations and trainable parameter counts. Each convex model and its corresponding conventional baseline use matched principal depth and width settings, while their parameter counts may differ because of architecture-specific constraints and parameter sharing.}
    \label{tab:model_hyperparameters}
    \footnotesize
    \setlength{\tabcolsep}{4pt}
    \renewcommand{\arraystretch}{1.15}
    \begin{tabular*}{\textwidth}{@{\extracolsep{\fill}}llccccrr@{}}
        \toprule
        \textbf{Model}
        & \textbf{Architecture}
        & \textbf{Layers}
        & \textbf{Model/hidden dim.}
        & \textbf{FF dim.}
        & \textbf{Heads}
        & \textbf{Apartment params}
        & \textbf{Office params} \\
        \midrule
        IC-EoT
        & Input convex attention
        & 1 & 64 & 128 & 1
        & 23,563 & 25,488 \\
        
        IC-LSTM
        & Input convex recurrent
        & 1 & 128 & -- & --
        & 22,511 & 29,436 \\
        
        EoT
        & Conventional attention
        & 1 & 64 & 128 & 1
        & 27,595 & 28,880 \\
        
        LSTM
        & Conventional recurrent
        & 1 & 128 & -- & --
        & 75,839 & 85,774 \\
        \bottomrule
    \end{tabular*}
\end{table*}

A historical sequence length of $N_h=10$ is used for the primary predictive-performance comparison. All four models are trained using Adam with a learning rate of $10^{-4}$, a batch size of 256, and a maximum of 2000 epochs. Early stopping is applied to the validation loss with a patience of 10 epochs, and the model weights corresponding to the lowest validation loss are restored. Within each testbed, all four models use the same training, validation, and test data.

\subsubsection{Predictive Performance and Training Cost}
\label{subsubsec:prediction_training_results}

Table~\ref{tab:prediction_training_performance} reports the predictive accuracy on standardised test outputs together with the mean time per training epoch. Mean squared error (MSE) and the coefficient of determination ($R^2$) are used as the principal accuracy metrics.

In addition to the average $R^2$ across all predicted variables, Table~\ref{tab:prediction_training_performance} reports the lowest $R^2$ among the controlled zone temperatures and the $R^2$ of facility electricity consumption. The former indicates whether the average score conceals a poorly predicted thermal zone, while the latter is relevant because predicted electricity consumption enters the MPC objective directly.

\begin{table*}[t]
    \centering
    \caption{Predictive performance and training time under the common $N_h=10$ configuration. MSE values are averaged across standardised output components. ``Worst-zone $R^2$'' denotes the minimum $R^2$ among the controlled zone temperatures.}
    \label{tab:prediction_training_performance}
    \footnotesize
    \setlength{\tabcolsep}{4pt}
    \renewcommand{\arraystretch}{1.15}
    \begin{tabular*}{\textwidth}{@{\extracolsep{\fill}}llccccc@{}}
        \toprule
        \textbf{Testbed}
        & \textbf{Model}
        & \textbf{Avg. MSE}
        & \textbf{Avg. $R^2$}
        & \textbf{Worst-zone $R^2$}
        & \textbf{Electricity $R^2$}
        & \textbf{Mean epoch time (s)} \\
        \midrule
        \multirow{4}{*}{ApartmentsThermal}
        & IC-EoT  & 0.0427 & 0.9357 & 0.9778 & 0.5918 & 0.5317 \\
        & IC-LSTM & 0.0379 & 0.9390 & 0.9722 & 0.6702 & 0.7201 \\
        & EoT     & 0.0196 & 0.9710 & 0.9881 & 0.8919 & 0.5989 \\
        & LSTM    & 0.0299 & 0.9614 & 0.9828 & 0.8038 & 1.1671 \\
        \midrule
        \multirow{4}{*}{OfficesThermostat}
        & IC-EoT  & 0.0332 & 0.9670 & 0.9139 & 0.8217 & 0.5243 \\
        & IC-LSTM & 0.0272 & 0.9729 & 0.9293 & 0.8499 & 0.7472 \\
        & EoT     & 0.0127 & 0.9875 & 0.9798 & 0.8693 & 0.5787 \\
        & LSTM    & 0.0134 & 0.9867 & 0.9796 & 0.8719 & 1.1480 \\
        \bottomrule
    \end{tabular*}
\end{table*}

IC-EoT and IC-LSTM achieve closely comparable overall predictive performance. Their average $R^2$ values differ by only 0.0033 in ApartmentsThermal and 0.0060 in OfficesThermostat. IC-LSTM provides a modestly better fit to facility electricity consumption, whereas IC-EoT achieves a slightly higher worst-zone temperature $R^2$ in ApartmentsThermal. Neither input convex architecture therefore consistently dominates the other in one-step predictive accuracy.

The conventional EoT and LSTM achieve lower MSE and higher average $R^2$ than their input convex counterparts. This performance gap is consistent with the representational restrictions imposed by input convexity, including non-negative weights, monotone convex compositions, and the removal or modification of operations that do not preserve convexity. The restrictions have a limited effect on most zone-temperature predictions but produce a more visible reduction in the electricity-consumption fit, particularly for ApartmentsThermal. The comparison illustrates the expected trade-off between unconstrained predictive flexibility and the convexity required for tractable MPC.

Under the common $N_h=10$ setting, IC-EoT records the lowest mean epoch time in both testbeds, requiring 0.5317~s per epoch in ApartmentsThermal and 0.5243~s in OfficesThermostat. Relative to IC-LSTM, these values correspond to reductions of approximately 26.2\% and 29.8\%, respectively. The difference between IC-EoT and the conventional EoT is smaller, indicating that the observed advantage mainly reflects the lower sequential computational burden relative to the recurrent architectures under the tested implementation.

\subsubsection{Numerical Training Stability of IC-EoT and IC-LSTM}
\label{subsubsec:numerical_training_stability}
The numerical training behaviour of IC-EoT and IC-LSTM is evaluated over a range of historical sequence lengths and random seeds. This design assesses whether the observed training outcomes are repeatable rather than being specific to a single training run.

IC-EoT and IC-LSTM are trained using five random seeds,
\begin{equation}
\mathcal{S}=\{42,43,44,45,46\},
\end{equation}
and six historical sequence lengths,
\begin{equation}
N_h\in\{5,10,15,20,25,30\}.
\end{equation}

For each model and testbed, combining the six historical sequence lengths with five random seeds gives 30 training runs. At every epoch, the training and validation losses and the occurrence of NaN or Inf values are recorded. An auxiliary safeguard terminates model fitting when both the training and validation losses remain NaN or Inf for three consecutive epochs. A non-finite value in only one of the two losses does not immediately terminate training, thereby allowing transient numerical events to recover.

Each run is assigned to one of four outcome classes:
\begin{itemize}
    \item \textbf{Stable (S):} no NaN or Inf value is detected in the training or validation loss throughout training;

    \item \textbf{Recovered (R):} a non-finite loss occurs during training, but the run completes and its final validation loss returns to within three times the median final validation loss of stable runs with the same testbed, model, and $N_h$;

    \item \textbf{Completed abnormal (A):} the training process completes, but its final validation loss remains non-finite or outside the above recovery range;

    \item \textbf{Interrupted (I):} persistent non-finite behaviour prevents a valid final evaluation from being obtained, including runs for which the safeguard terminates model fitting after both losses remain non-finite for three consecutive epochs.
\end{itemize}
The numerically successful rate is defined as the proportion of Stable and Recovered runs. Table~\ref{tab:multiseed_stability} reports both the successful rate and the complete S/R/A/I outcome distribution for every model, testbed, and historical sequence length.

\begin{table*}[t]
    \centering
    \caption{Multi-seed numerical training outcomes for the two input convex models. Each model--$N_h$ configuration contains five training runs with distinct random seeds. Outcome counts are reported as S/R/A/I, denoting Stable, Recovered, Completed abnormal, and Interrupted runs, respectively.}
    \label{tab:multiseed_stability}
    \footnotesize
    \setlength{\tabcolsep}{4pt}
    \renewcommand{\arraystretch}{1.15}
    \begin{tabular*}{\textwidth}{@{\extracolsep{\fill}}lllcccccc@{}}
        \toprule
        \textbf{Testbed}
        & \textbf{Model}
        & \textbf{Statistic}
        & \multicolumn{6}{c}{\textbf{Historical sequence length} $\boldsymbol{N_h}$} \\
        \cmidrule(lr){4-9}
        &
        &
        & $\boldsymbol{5}$
        & $\boldsymbol{10}$
        & $\boldsymbol{15}$
        & $\boldsymbol{20}$
        & $\boldsymbol{25}$
        & $\boldsymbol{30}$ \\
        \midrule

        \multirow{4}{*}{ApartmentsThermal}
        & \multirow{2}{*}{IC-EoT}
        & Successful rate
        & 100\% & 100\% & 100\% & 100\% & 100\% & 100\% \\
        &
        & S/R/A/I
        & 5/0/0/0 & 5/0/0/0 & 5/0/0/0
        & 5/0/0/0 & 5/0/0/0 & 5/0/0/0 \\
        \cmidrule(lr){2-9}
        & \multirow{2}{*}{IC-LSTM}
        & Successful rate
        & 100\% & 100\% & 100\% & 80\% & 60\% & 80\% \\
        &
        & S/R/A/I
        & 5/0/0/0 & 5/0/0/0 & 5/0/0/0
        & 4/0/0/1 & 3/0/1/1 & 4/0/0/1 \\

        \midrule

        \multirow{4}{*}{OfficesThermostat}
        & \multirow{2}{*}{IC-EoT}
        & Successful rate
        & 100\% & 100\% & 100\% & 100\% & 100\% & 100\% \\
        &
        & S/R/A/I
        & 5/0/0/0 & 5/0/0/0 & 5/0/0/0
        & 5/0/0/0 & 5/0/0/0 & 5/0/0/0 \\
        \cmidrule(lr){2-9}
        & \multirow{2}{*}{IC-LSTM}
        & Successful rate
        & 100\% & 100\% & 80\% & 80\% & 100\% & 60\% \\
        &
        & S/R/A/I
        & 5/0/0/0 & 5/0/0/0 & 4/0/1/0
        & 3/1/0/1 & 4/1/0/0 & 2/1/1/1 \\

        \bottomrule
    \end{tabular*}
\end{table*}

As shown in Table~\ref{tab:multiseed_stability}, IC-EoT remains stable in all 60 runs across the two testbeds. No NaN or Inf value is detected for any tested random seed or sequence length. IC-LSTM is stable in all runs for $N_h=5$ and $N_h=10$, but non-finite events emerge as the historical sequence is lengthened.

In ApartmentsThermal, all IC-LSTM runs remain stable up to $N_h=15$. The successful rate subsequently decreases to 80\% at $N_h=20$, 60\% at $N_h=25$, and 80\% at $N_h=30$. Across all 30 Apartment runs, 26 are classified as Stable, one as Completed abnormal, and three as Interrupted. In OfficesThermostat, the successful rates are 100\%, 100\%, 80\%, 80\%, 100\%, and 60\% for $N_h=5,10,15,20,25,$ and 30, respectively. Across all 30 Office runs, 23 are classified as Stable, three as Recovered, two as Completed abnormal, and two as Interrupted. The outcome is not strictly monotonic in $N_h$, indicating that the random seed and the resulting optimization  trajectory also influence whether numerical instability is triggered. Nevertheless, every anomalous Apartment run occurs at $N_h\geq20$, and every anomalous Office run occurs at $N_h\geq15$.

\begin{figure*}[t]
    \centering
    \includegraphics[width=0.82\textwidth]
    {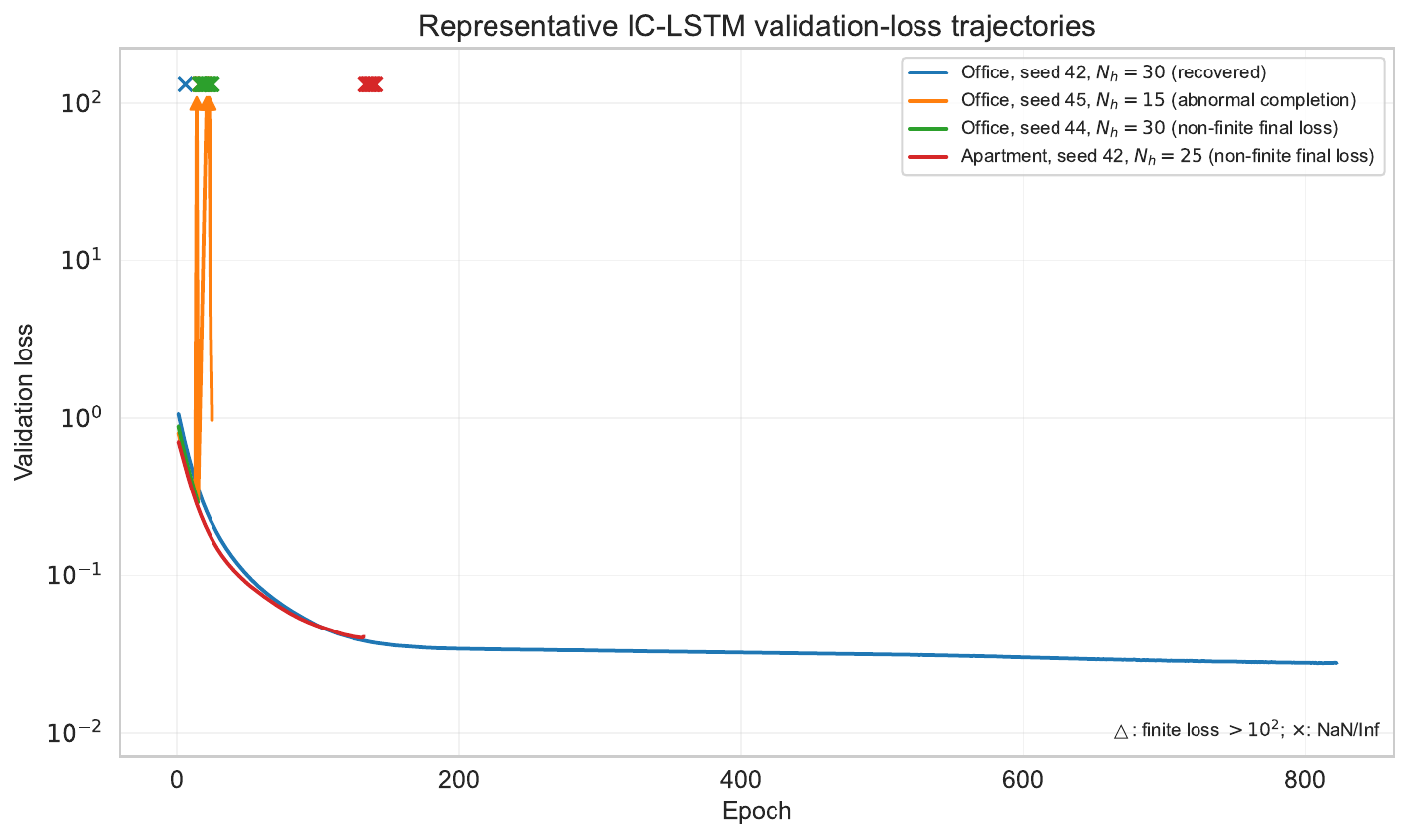}
    \caption{Representative IC-LSTM validation-loss trajectories. Finite validation losses above $10^2$ are clipped to the upper plotting boundary and marked by triangles, while NaN/Inf values are marked by crosses. The selected runs illustrate recovered, abnormal-completion, and non-finite-final-loss outcomes.}
    \label{fig:representative_instability_trajectories}
\end{figure*}

Figure~\ref{fig:representative_instability_trajectories} illustrates the distinction among the outcome classes. For OfficesThermostat with seed 42 and $N_h=30$, the validation loss becomes NaN at an early epoch but immediately recovers; the run subsequently completes 822 epochs with a final validation loss of approximately 0.0275 and normal test accuracy. In contrast, the Office run with seed 45 and $N_h=15$ experiences NaN values and finite loss excursions spanning many orders of magnitude, terminates after 25 epochs with a final validation loss of approximately 0.964, and produces substantially degraded predictions. The Office run with seed 44 and $N_h=30$, and the Apartment run with seed 42 and $N_h=25$, both finish with non-finite validation losses.

The representative trajectories show that the non-finite events do not lead to a single uniform outcome. Some runs recover and subsequently converge to the normal validation-loss range, whereas others complete with abnormal losses or fail to produce a finite final evaluation. The observed phenomenon is therefore characterised as sequence-length-dependent numerical training instability.

The behaviour is nevertheless consistent with the IC-LSTM structure. Its state update contains repeated element-wise products between the forget gate and the previous cell state, and between the input gate and the candidate state. Because the convexity-compatible gates are non-negative and unbounded, gate values greater than one can repeatedly amplify internal states. Non-negative weight constraints also reduce cancellation between positive and negative contributions. As $N_h$ increases, both the forward computation and back-propagation traverse a longer sequence of multiplicative recurrent operations, increasing the opportunity for floating-point overflow or non-finite intermediate values.

IC-EoT contains no recurrent cell state and therefore avoids this specific chain of repeated multiplicative state transitions. Its historical inputs are processed through a finite stack of attention and feed-forward operations, so the depth of its temporal computation is determined by the encoder stack rather than by a recurrent path whose unrolled length grows with $N_h$. The experiment does not establish that IC-EoT is immune to every possible form of neural-network training instability. It shows that no non-finite event is observed across the tested sequence lengths, random seeds, and building datasets, 
while the recurrent input convex counterpart exhibits repeated numerical failures at longer historical sequence lengths.

\subsection{MPC Performance Evaluation}
\label{subsec:mpc_performance_evaluation}

This subsection evaluates the four predictive models when embedded in the receding-horizon MPC formulation introduced in Section~\ref{subsec:mpc_optimization_problem}. The comparison considers online solution time, numerical termination behaviour, electricity cost, and thermal-comfort performance. Prediction horizons of
\begin{equation}
N_p \in \{4,8,12,16,20,24,28,32\}
\end{equation}
are evaluated. With a 15-minute control interval, these values correspond to look-ahead periods ranging from one to eight hours.

\subsubsection{Evaluation Scenarios and Time-of-Use Tariffs}
\label{subsubsec:mpc_scenarios_tariffs}

Both testbeds are evaluated under summer operating conditions. The OfficesThermostat environment contains active mechanical cooling and is therefore used as the principal closed-loop cooling-control testbed. By contrast, ApartmentsThermal does not provide an actively controllable cooling source. The apartment MPC can adjust thermostat setpoints, but it cannot directly command mechanical cooling. Its cost and comfort results are therefore reported as reference results under limited cooling authority, while this testbed is used primarily to provide an additional assessment of online solution time.

For ApartmentsThermal, the Spanish residential time-of-use tariff adopted from Iberdrola Espa\~na's Plan Ahorro Solar is used~\cite{iberdrola_ahorro_solar}:
\begin{itemize}
    \item \textbf{Off-Peak / Valle (00:00--08:00):} \texteuro0.089/kWh on weekdays, with the off-peak rate applied throughout weekends and public holidays.
    \item \textbf{Mid-Peak / Llano (08:00--10:00, 14:00--18:00, and 22:00--24:00):} \texteuro0.135/kWh on weekdays.
    \item \textbf{Peak / Punta (10:00--14:00 and 18:00--22:00):} \texteuro0.210/kWh on weekdays.
\end{itemize}
This structure incentivises shifting flexible electricity use from the peak periods towards lower-price periods.

For OfficesThermostat, the PPC $\Gamma 23$ Business tariff is adopted, with the electricity charges taken from the official PPC tariff schedule~\cite{PPC_G23_2026}. The corresponding summer dual-zone periods are specified by the Greek distribution-network operator HEDNO~\cite{HEDNO_DualZone_2025}:
\begin{itemize}
    \item \textbf{Reduced Rate (02:00--04:00 and 11:00--15:00):} \texteuro0.129/kWh.
    \item \textbf{Regular Rate (all remaining hours):} \texteuro0.209/kWh.
\end{itemize}
The monthly fixed charge is excluded because it is independent of the control actions and identical for all evaluated controllers.

\subsubsection{Solver Time and Termination Behaviour}
\label{subsubsec:solver_time_reliability}

\begin{figure*}[t]
    \centering
    \begin{minipage}[t]{0.49\textwidth}
        \centering
        \includegraphics[width=\linewidth]{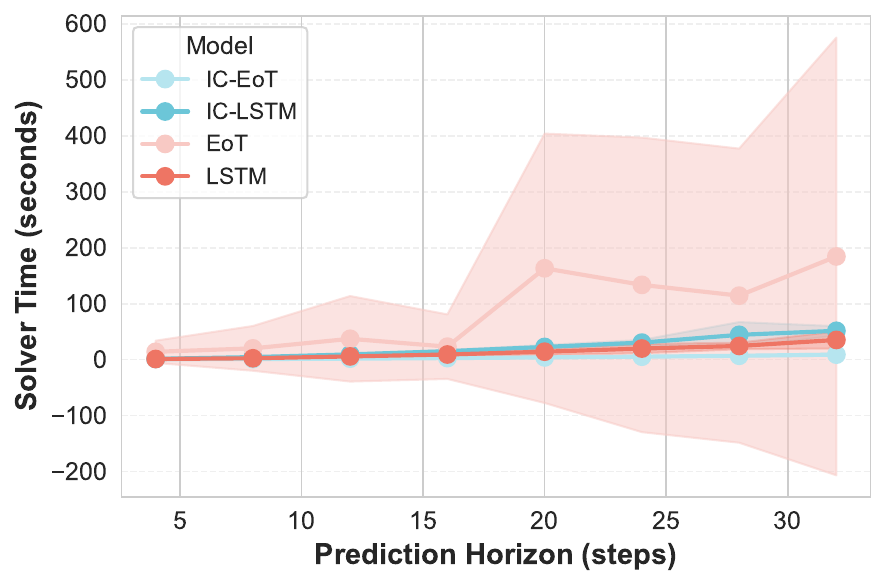}
        \par\small (a) ApartmentsThermal
    \end{minipage}
    \hfill
    \begin{minipage}[t]{0.49\textwidth}
        \centering
        \includegraphics[width=\linewidth]{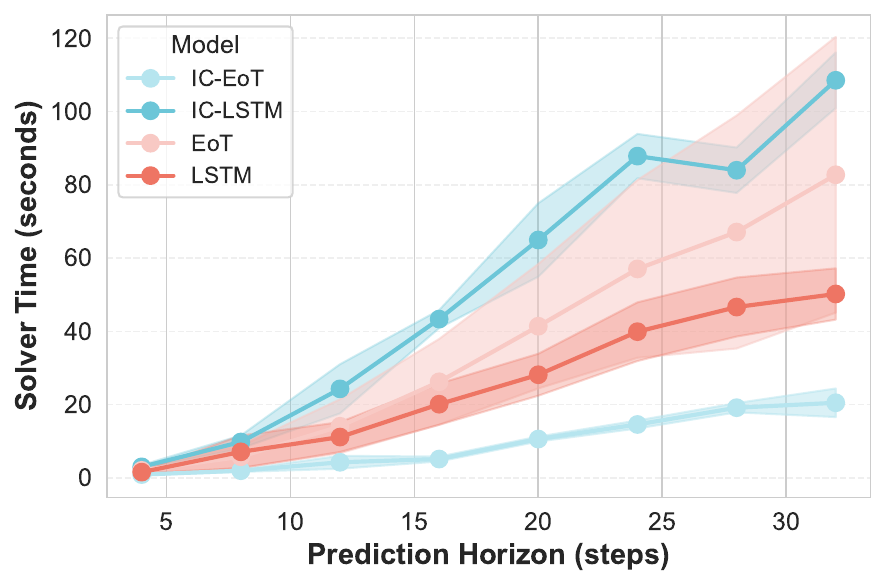}
        \par\small (b) OfficesThermostat
    \end{minipage}
    \caption{MPC solution time as a function of the prediction horizon. Markers show the mean wall-clock time per control step and shaded regions show one standard deviation over the closed-loop simulation.}
    \label{fig:solver_time_both_testbeds}
\end{figure*}

All model--horizon combinations are solved with IPOPT using a maximum of 300 iterations per MPC call. The same solver configuration is used for all four models within each testbed. Figure~\ref{fig:solver_time_both_testbeds} compares the wall-clock solution time per control step. Each marker denotes the mean over the closed-loop simulation, and the shaded region denotes one standard deviation.

In ApartmentsThermal, the mean solution time of IC-EoT increases from 0.990~s at $N_p=4$ to 8.883~s at $N_p=32$. At the longest horizon, IC-LSTM, LSTM, and EoT require 51.500~s, 35.157~s, and 184.394~s, respectively. IC-EoT is therefore approximately $5.8\times$, $4.0\times$, and $20.8\times$ faster than IC-LSTM, LSTM, and EoT, respectively, at $N_p=32$.

A similar advantage is observed in OfficesThermostat. The mean IC-EoT solution time increases from 0.920~s at $N_p=4$ to 20.503~s at $N_p=32$, whereas IC-LSTM, LSTM, and EoT require 108.497~s, 50.188~s, and 82.713~s at $N_p=32$. The corresponding IC-EoT speed-ups are approximately $5.3\times$, $2.4\times$, and $4.0\times$.

The solver-time results show that IC-EoT scales more favourably than the recurrent input convex baseline as the prediction horizon increases. IC-LSTM propagates recurrent hidden and cell states across the complete historical window during each model and derivative evaluation, whereas IC-EoT processes the same window through a fixed-depth encoder. This structural difference reduces the computational burden of the repeated model evaluations performed within IPOPT. The reported wall-clock times nevertheless reflect the complete MPC solve, including model evaluation, derivative computation, solver iterations, and termination behaviour.

The reported solution-time statistics retain all MPC calls, including calls that terminate before satisfying either convergence criterion, and therefore represent the end-to-end computational burden of the implemented controllers. Table~\ref{tab:solver_status_all_experiments} reports the IPOPT termination distribution for every testbed, model, and prediction horizon. Each cell follows the order S/A/M, where S denotes termination at the desired tolerances, A denotes termination at the acceptable tolerances, and M denotes termination at the 300-iteration limit. Each model--horizon experiment contains 86 MPC solves. No other IPOPT termination status is observed.

\begin{table*}[t]
    \centering
    \caption{IPOPT termination-status counts for all MPC experiments. Each horizon cell reports S/A/M, denoting desired-tolerance, acceptable-tolerance, and maximum-iteration terminations, respectively. The final column gives the valid-termination rate, $(\mathrm{S}+\mathrm{A})/\mathrm{Total}$, across all eight horizons.}
    \label{tab:solver_status_all_experiments}
    \scriptsize
    \setlength{\tabcolsep}{3.0pt}
    \renewcommand{\arraystretch}{1.13}
    \resizebox{\textwidth}{!}{%
    \begin{tabular}{llccccccccc}
        \toprule
        \textbf{Testbed}
        & \textbf{Model}
        & $\boldsymbol{N_p=4}$
        & $\boldsymbol{N_p=8}$
        & $\boldsymbol{N_p=12}$
        & $\boldsymbol{N_p=16}$
        & $\boldsymbol{N_p=20}$
        & $\boldsymbol{N_p=24}$
        & $\boldsymbol{N_p=28}$
        & $\boldsymbol{N_p=32}$
        & \textbf{Valid} \\
        \midrule

        \multirow{4}{*}{ApartmentsThermal}
        & IC-EoT
        & 85/0/1
        & 86/0/0
        & 86/0/0
        & 86/0/0
        & 86/0/0
        & 86/0/0
        & 86/0/0
        & 86/0/0
        & 99.9\% \\

        & IC-LSTM
        & 86/0/0
        & 86/0/0
        & 86/0/0
        & 86/0/0
        & 86/0/0
        & 86/0/0
        & 86/0/0
        & 86/0/0
        & 100.0\% \\

        & EoT
        & 61/0/25
        & 74/0/12
        & 74/0/12
        & 84/0/2
        & 65/0/21
        & 74/0/12
        & 80/0/6
        & 76/0/10
        & 85.5\% \\

        & LSTM
        & 86/0/0
        & 86/0/0
        & 86/0/0
        & 86/0/0
        & 86/0/0
        & 86/0/0
        & 86/0/0
        & 86/0/0
        & 100.0\% \\

        \midrule

        \multirow{4}{*}{OfficesThermostat}
        & IC-EoT
        & 86/0/0
        & 86/0/0
        & 85/1/0
        & 86/0/0
        & 86/0/0
        & 86/0/0
        & 86/0/0
        & 86/0/0
        & 100.0\% \\

        & IC-LSTM
        & 86/0/0
        & 86/0/0
        & 86/0/0
        & 86/0/0
        & 86/0/0
        & 86/0/0
        & 86/0/0
        & 86/0/0
        & 100.0\% \\

        & EoT
        & 81/5/0
        & 79/7/0
        & 62/24/0
        & 51/35/0
        & 46/40/0
        & 47/39/0
        & 53/33/0
        & 59/27/0
        & 100.0\% \\

        & LSTM
        & 86/0/0
        & 81/5/0
        & 84/2/0
        & 86/0/0
        & 86/0/0
        & 85/1/0
        & 86/0/0
        & 86/0/0
        & 100.0\% \\

        \bottomrule
    \end{tabular}%
    }
\end{table*}

The two input convex models exhibit highly consistent termination behaviour across the tested configurations. IC-LSTM reaches the desired tolerances in all 1376 MPC calls across the two testbeds. IC-EoT reaches the desired tolerances in 1374 calls, terminates at the acceptable tolerance once in OfficesThermostat at $N_p=12$, and reaches the 300-iteration limit once in ApartmentsThermal at $N_p=4$. Its corresponding valid-termination rates are therefore 99.9\% and 100.0\% in the apartment and office testbeds, respectively.

The two isolated IC-EoT outcomes reflect the finite numerical tolerances and iteration budget used by IPOPT. Convexity ensures that a converged local optimum is globally optimal for the corresponding learned formulation, but it does not prescribe the number of iterations required to satisfy a particular numerical stopping criterion. The acceptable-level termination is counted as valid, whereas the iteration-limit call is retained in the end-to-end solver-time statistics.

The two conventional models exhibit different termination behaviour in ApartmentsThermal. EoT reaches the 300-iteration limit in 100 of the 688 MPC calls, giving an overall valid-termination rate of 85.5\%. Iteration-limit events occur at every tested horizon, with the largest counts observed at $N_p=4$ and $N_p=20$. These calls contribute to the substantially larger mean and variance of the EoT solver time, particularly at the longer horizons. By contrast, LSTM reaches the desired tolerances in all 688 apartment MPC calls.

In OfficesThermostat, all four models obtain a 100\% valid-termination rate. IC-LSTM reaches the desired tolerances in all 688 calls, while IC-EoT records 687 desired-tolerance terminations and one acceptable-level termination. EoT terminates at the acceptable level in 210 calls, compared with eight such calls for LSTM. Although these outcomes satisfy the adopted valid-termination criterion, their frequency shows that EoT more often meets the relaxed acceptable conditions rather than the desired tolerances.

Overall, the results provide a global-optimality interpretation for converged IC-EoT- and IC-LSTM-based MPC problems, while the same guarantee is unavailable for the conventional EoT and LSTM formulations. The results do not imply that convexity alone determines wall-clock time or solver termination. Model-evaluation cost, derivative computation, numerical scaling, active constraints, initialisation, and the imposed iteration budget also affect the observed solver behaviour.

\subsubsection{Closed-Loop Cost and Comfort Performance}
\label{subsubsec:closed_loop_cost_comfort}

Closed-loop performance is evaluated using the total electricity bill and the average upper-temperature-violation degree-hours. Degree-hours account for both the magnitude and duration of overheating and are averaged across the controlled zones. The statistics use a common full-day accounting boundary for all models and horizons.

\begin{figure*}[t]
    \centering
    \begin{minipage}[t]{0.49\textwidth}
        \centering
        \includegraphics[width=\linewidth]{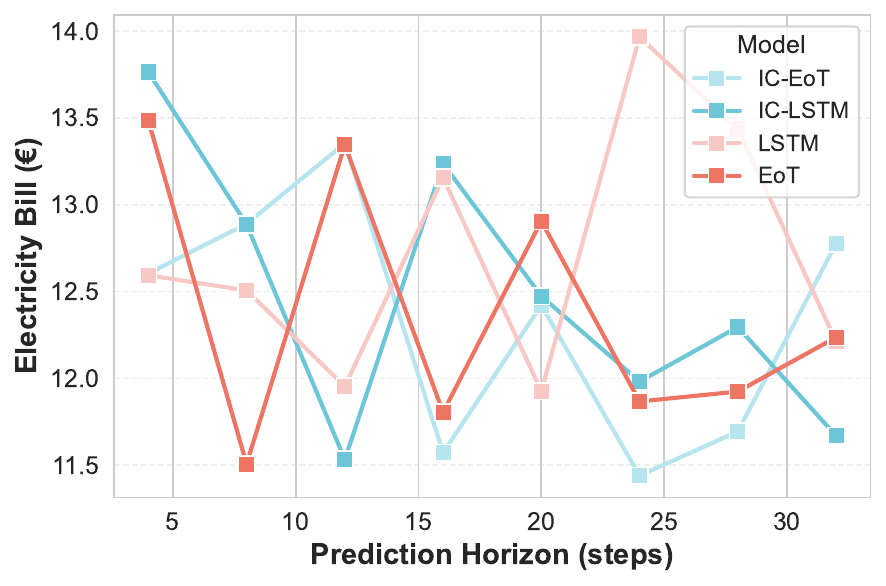}
        \par\small (a) Apartment electricity bill
    \end{minipage}
    \hfill
    \begin{minipage}[t]{0.49\textwidth}
        \centering
        \includegraphics[width=\linewidth]{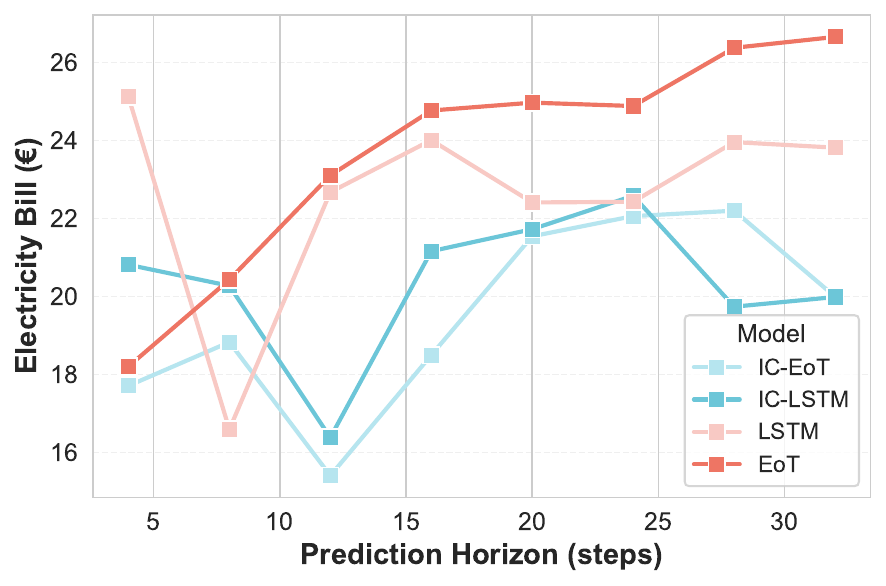}
        \par\small (b) Office electricity bill
    \end{minipage}

    \vspace{2mm}

    \begin{minipage}[t]{0.49\textwidth}
        \centering
        \includegraphics[width=\linewidth]{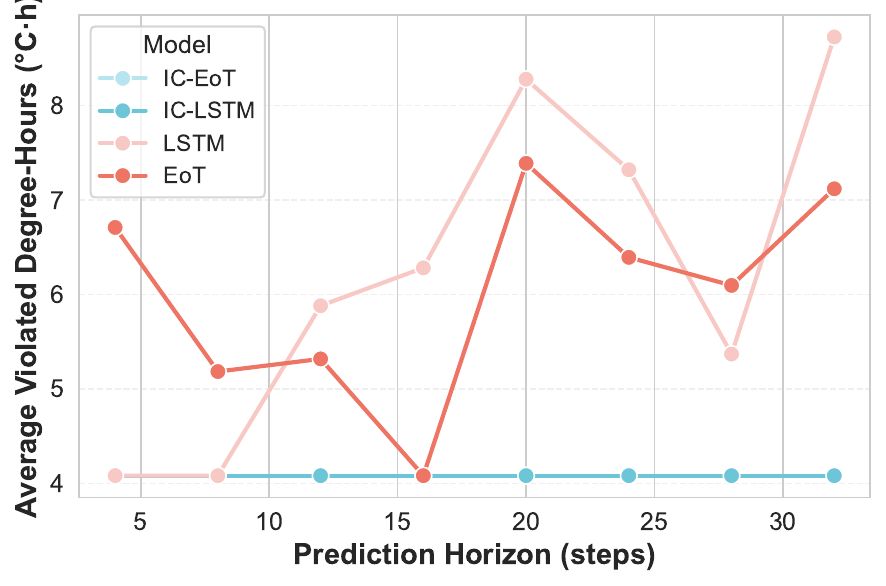}
        \par\small (c) Apartment comfort violation
    \end{minipage}
    \hfill
    \begin{minipage}[t]{0.49\textwidth}
        \centering
        \includegraphics[width=\linewidth]{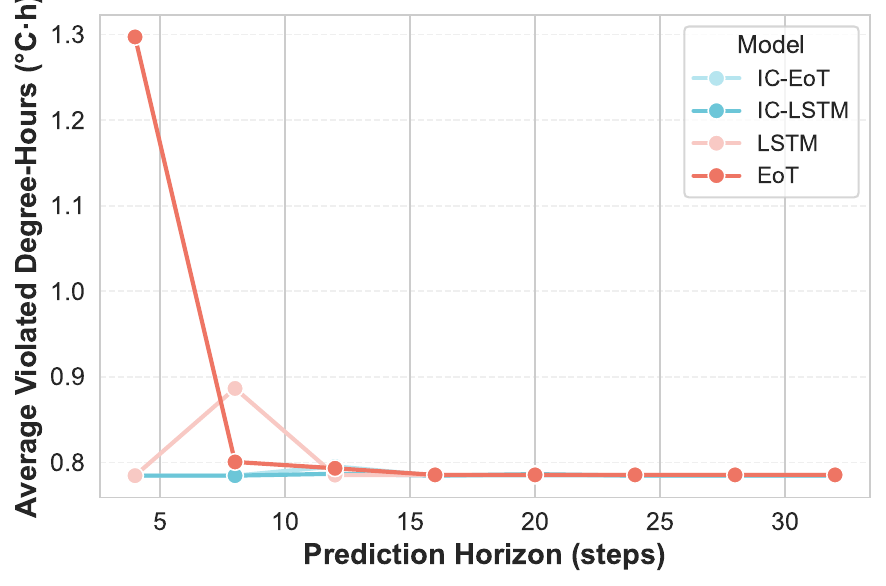}
        \par\small (d) Office comfort violation
    \end{minipage}

    \caption{Closed-loop electricity cost and thermal-comfort performance as functions of the prediction horizon. Panels (a) and (b) report the full-day electricity bill, while panels (c) and (d) report the average upper-temperature-violation degree-hours across the controlled zones.}
    \label{fig:cost_comfort_both_testbeds}
\end{figure*}

\begin{table*}[t]
    \centering
    \caption{Summary of full-day closed-loop cost and comfort performance across the tested prediction horizons. The minimum electricity bill is accompanied by the corresponding prediction horizon.}
    \label{tab:cost_comfort_summary}
    \small
    \setlength{\tabcolsep}{6pt}
    \renewcommand{\arraystretch}{1.15}
    \begin{tabular}{llcccc}
        \toprule
        \textbf{Testbed} & \textbf{Model} & \textbf{Minimum bill (\texteuro)} & \textbf{Mean bill (\texteuro)} & \textbf{Mean DH ($^{\circ}$C$\cdot$h)} & \textbf{Maximum DH ($^{\circ}$C$\cdot$h)} \\
        \midrule
        \multirow{4}{*}{ApartmentsThermal}
        & IC-EoT  & 11.44 ($N_p=24$) & 12.34 & 4.0826 & 4.0826 \\
        & IC-LSTM & 11.53 ($N_p=12$) & 12.48 & 4.0826 & 4.0826 \\
        & EoT     & 11.51 ($N_p=8$)  & 12.39 & 6.0366 & 7.3886 \\
        & LSTM    & 11.93 ($N_p=20$) & 12.72 & 6.2531 & 8.7266 \\
        \midrule
        \multirow{4}{*}{OfficesThermostat}
        & IC-EoT  & 15.40 ($N_p=12$) & 19.53 & 0.7859 & 0.7954 \\
        & IC-LSTM & 16.38 ($N_p=12$) & 20.33 & 0.7850 & 0.7867 \\
        & EoT     & 18.20 ($N_p=4$)  & 23.67 & 0.8522 & 1.2973 \\
        & LSTM    & 16.59 ($N_p=8$)  & 22.63 & 0.7978 & 0.8864 \\
        \bottomrule
    \end{tabular}
\end{table*}

For ApartmentsThermal, IC-EoT and IC-LSTM produce identical mean degree-hours of $4.0826~^{\circ}\mathrm{C}\cdot\mathrm{h}$ at every tested horizon. Their mean electricity bills are also similar, at \texteuro12.34 and \texteuro12.48, respectively. IC-EoT obtains both the lowest mean bill and the lowest individual bill among the four controllers, with a minimum of \texteuro11.44 at $N_p=24$.

The two non-convex baselines exhibit greater horizon-dependent variation in thermal-comfort performance. EoT records mean and maximum degree-hours of 6.0366 and $7.3886~^{\circ}\mathrm{C}\cdot\mathrm{h}$, respectively, while the corresponding LSTM values are 6.2531 and $8.7266~^{\circ}\mathrm{C}\cdot\mathrm{h}$. Their mean electricity bills are \texteuro12.39 and \texteuro12.72. Although the apartment environment provides only limited cooling authority, the input convex controllers produce more consistent cost and comfort outcomes across the tested horizons. These results should be interpreted as a relative comparison among the four learned controllers rather than as evidence of full active cooling capability.

OfficesThermostat provides the more representative active-cooling comparison. IC-EoT and IC-LSTM achieve almost identical comfort performance, with mean degree-hours of 0.7859 and $0.7850~^{\circ}\mathrm{C}\cdot\mathrm{h}$, respectively. IC-EoT obtains a lower mean electricity bill, \texteuro19.53 compared with \texteuro20.33 for IC-LSTM, while requiring substantially less online solution time. EoT and LSTM exhibit higher mean bills and greater horizon-dependent variation in comfort performance.

These closed-loop results should not be interpreted as a direct consequence of convexity alone. Convexity concerns optimality with respect to the learned predictive model. The realised EnergyPlus performance also depends on prediction error, model--plant mismatch, disturbance forecasts, receding-horizon effects, and available physical actuation. A non-convex surrogate may therefore occasionally produce a lower realised bill at an individual horizon, while a globally optimal solution of a convex surrogate need not be globally optimal for the true building. Overall, IC-EoT provides cost and comfort performance comparable to IC-LSTM while achieving substantially more favourable online solution-time scaling.

\section{Conclusion and Future Work}
\label{sec:conclusion_future_work}

This paper proposed the IC-EoT for learning-based building MPC. The architecture replaces recurrent state propagation with an encoder-only structure and employs convex additive attention, non-negative affine mappings, convex nondecreasing activations and residual additions. Theoretical analysis showed that the proposed predictor is input convex and that, under the stated assumptions, recursive multi-step prediction preserves the convexity required by the soft-constrained MPC formulation.

The model was evaluated in the residential ApartmentsThermal and commercial OfficesThermostat testbeds against IC-LSTM, EoT, and LSTM. IC-EoT achieved predictive accuracy close to that of IC-LSTM while reducing the mean training time per epoch by 26.2\% and 29.8\% in the two testbeds, respectively. Across 60 multi-seed runs covering historical sequence lengths from 5 to 30, no non-finite training event was observed for IC-EoT, whereas IC-LSTM exhibited repeated numerical anomalies at longer sequence lengths. In MPC, IC-EoT achieved valid-termination rates of 99.9\% and 100.0\%. At the eight-hour prediction horizon, its mean solution time was approximately $5.8\times$ and $5.3\times$ lower than that of IC-LSTM in ApartmentsThermal and OfficesThermostat, respectively, while providing comparable closed-loop cost and comfort performance. These findings demonstrate an empirical computational and numerical advantage over the recurrent input convex baseline, rather than a general guarantee of training stability or universally superior control performance.

The current results are limited to two EnergyPlus-based testbeds and selected summer operating conditions. In particular, ApartmentsThermal does not provide active mechanical cooling, so its closed-loop results mainly serve as a supplementary computational comparison. Future work will evaluate IC-EoT over longer operating periods, additional buildings, and physical or hardware-in-the-loop systems. Further research will also investigate robust or stochastic MPC formulations, uncertainty-aware constraint handling, and more expressive convex attention architectures to reduce the predictive-accuracy gap while retaining convexity and computational efficiency.

\section*{CRediT authorship contribution statement}

Kaipeng Xu: Conceptualization, Methodology, Software, Validation, Formal analysis, Investigation, Data curation, Visualization, Writing -- original draft, Writing -- review \& editing, Project administration. Zhuo Zhi: Software, Writing -- review \& editing. Keyue Jiang: Writing -- review \& editing.

\section*{Declaration of competing interest}

The authors declare that they have no known competing financial interests or personal relationships that could have appeared to influence the work reported in this paper.

\section*{Data availability}

Data and code supporting the findings of this study will be made available on reasonable request.

\bibliographystyle{elsarticle-num}
\bibliography{literature}

\end{document}